\documentclass[usenatbib]{mnras}

\usepackage[T1]{fontenc}
\usepackage{ae,aecompl}

\usepackage{graphicx}	% Including figure files
\usepackage{amsmath}	% Advanced maths commands
\usepackage{amssymb}	% Extra maths symbols
\usepackage{makecell} 
\usepackage{bm}
\newcommand{\gradZ}{ \nabla Z }
\newcommand{\Zchar}{ Z_{\text{char}} }
\newcommand{\Hii}{H\textsc{ii} }

\newcommand{\PHX}{\textsc{pyHIIextractor}}
\newcommand{\HiiPhot}{\textsc{HIIphot}}
\newcommand{\NSH}{N$_2$S$_2$H$\alpha$}

\newcommand{\ON}{O$_3$N$_2$}
\newcommand{\RS}{RS$_{32}$}

\defcitealias{Geogals1}{M21}
\newcommand{\GeoGalsI}{\citetalias{Geogals1}}
\defcitealias{Geogals2}{M22}
\newcommand{\GeoGalsII}{\citetalias{Geogals2}}
\title[Geostatistics of Galaxies III]{A geostatistical analysis of multiscale metallicity variations in galaxies - III. Spatial resolution and data quality limits}

\author[Metha, B. et al.]{
Benjamin Metha$^{1,2,3}$\thanks{\hbox{methab@student.unimelb.edu.au}}, Michele Trenti$^{1,2}$, Andrew Battisti$^{2,5}$, Tingjin Chu$^{4}$ %, Tommaso Treu$^3$
\\
$^1$School of Physics, The University of Melbourne, VIC 3010, Australia\\
$^2$Australian Research Council Centre of Excellence for All-Sky Astrophysics in 3-Dimensions, Australia \\
$^3$Department of Physics and Astronomy, University of California, Los Angeles, 430 Portola Plaza, Los Angeles, CA 90095, USA\\
$^4$School of Mathematics and Statistics, The University of Melbourne, VIC 3010, Australia\\
$^5$Research School of Astronomy and Astrophysics, Australian National University, Cotter Road, Weston Creek, ACT 2611, Australia
}

\date{Accepted XXX. Received YYY; in original form ZZZ}

\pubyear{2023}

\begin{document}
\label{firstpage}
\pagerange{\pageref{firstpage}--\pageref{lastpage}}
    \maketitle

\begin{abstract}
Geostatistical methods are powerful tools for understanding the spatial structure of the metallicity distribution of galaxies, and enable construction of accurate predictive models of the 2D metallicity distribution. However, so far these methods have only been applied to very high spatial resolution metallicity maps, leaving it uncertain if they will work on lower quality data. In this study, we apply geostatistical techniques to high-resolution spectroscopic maps of three local galaxies convolved to eight different spatial resolutions ranging from $\sim 40$pc to $\sim 1$ kpc per pixel. We fit a geostatistical model to the data at all resolutions, and find that
for metallicity maps where small scale structure is visible by eye (with $\gtrsim 10$ resolution elements per $R_e$), all parameters, including the metallicity correlation scale, can be recovered accurately. At all resolutions tested, we find that point metallicity predictions from such a geostatistical model outperform a circularly symmetric metallicity gradient model. We also explore dependence on the number of data points, and find that $N\gtrsim 100$ spatially resolved metallicity values are sufficient to train a geostatistical model that yields more accurate metallicity predictions than a radial gradient model. Finally, we investigate the potential detrimental effects of having spaxels smaller than an individual \Hii region by repeating our analysis with metallicities integrated over \Hii regions. We see that spaxel-based measurements have more noise, as expected, but the underlying spatial metallicity distribution can be recovered regardless of whether spaxels or integrated regions are used.
\end{abstract}


%%%%%%%%%%%%%%%%%%%%%%%%%%%%%%%%%%%%%%%%%%%%%%%%%%%%
%%%%%%%%%%%%%%%%%%%%%%%%%%%%%%%%%%%%%%%%%%%%%%%%%%%%
%%%%%%%%%%%%%%%%% BODY OF PAPER %%%%%%%%%%%%%%%%%%%%
%%%%%%%%%%%%%%%%%%%%%%%%%%%%%%%%%%%%%%%%%%%%%%%%%%%%
%%%%%%%%%%%%%%%%%%%%%%%%%%%%%%%%%%%%%%%%%%%%%%%%%%%%

\section{Introduction} \label{sec:intro}

The distribution of metals (chemical elements heavier than Helium) throughout a galaxy is sensitive to many details of their formation and evolution that are not yet fully understood. It is well established that metals are released into the interstellar medium (ISM) of galaxies by supernovae \citep{Andrews+20}, but we do not know how efficiently they can mix into their local environment \citep{Li+21}. We know that a fraction of the metals released in these supernovae ejecta escape the galaxy via metal enriched outflows \citep[e.g][]{Heckman+90, Reichardt-Chu+22}, but we do not know how large this escape fraction is, nor how much of the ejected material falls back onto the galaxy later in a galactic fountain \citep{Christensen+18}. We know that galaxies accrete metal-poor gas to fuel star formation \citep{Larson72, Trapp+22}, but there is still some debate as to whether this happens preferentially on the central regions or the outskirts of galactic discs \citep{Schonrich+McMillan17}.
Detailed analysis of the internal chemical structure of galaxies can shed light on the relative strengths and timescales of all of these processes, making it a key element for addressing many theoretical challenges in understanding galaxy formation and evolution \citep{Naab+17}.

Traditionally, the internal metallicity structure of galaxies has been understood through the lens of a radial linear gradient model \citep{Searle71}. Through a large collection of observational campaigns  \citep[e.g.][to name a few]{VilaCostas+92, Zaritsky+94, Croom+12, CALIFA, Berg+13, Ho+15, SAMI, MANGA, Belfiore+2017}, it has been established that most spiral galaxies in the local universe follow a negative metallicity gradient, with more metal enrichment in their central regions than in the outskirts of their disks. This can be explained through an ``inside-out" galaxy formation model, in which star formation begins first in the central regions of a galaxy, allowing those regions to be more metal-enriched through successive generations of star formation than the outskirts, where star-formation has only recently begun \citep[e.g][]{Boissier+Prantzos99}.

Recent technological advancements in high resolution integral field spectroscopy (IFS) has led to acquisition of detailed metallicity maps for galaxies in the local universe, with resolutions finer than $\sim 100$ parsec \citep{Erroz-Ferrer+19, Lopez-Coba+20, Emsellem+22}. Using data from MUSE, and psuedo-IFS data obtained through the TYPHOON survey (Seibert et al. in prep.), large azimuthal variations in metallicity have been seen to exist in many local spiral galaxies in addition to a metallicity gradient \citep[e.g.][]{Sanchez-Menguiano+16, Sanchez-Menguiano+18, Ho+17, Ho+18, Ho+19, Kreckel+19}. This data has also been used to show that small-scale variations in metallicity exist in addition to a radial trend, and to analyse the spatial scale of these variations, using a variety of statistical techniques \citep{Kreckel+20, Li+21, Li+22, Williams+22}.

In the first two papers of this series (\citealt{Geogals1}, hereafter \GeoGalsI; and \citealt{Geogals2}, hereafter \GeoGalsII), we have explored how tools and techniques from geostatistics, the mathematical study of stochastic processes that occur over a continuous spatial domain, can be applied to analyse data products from high-resolution studies of metallicity variations within star-forming galaxies in the local universe. In \GeoGalsI, seven galaxies from the PHANGS-MUSE survey were analysed using a geostatistical approach to recover the statistical properties of their small-scale metallicity fluctuations, allowing predictions of an analytical model to be tested. In \GeoGalsII, geostatistical models were fit to high-resolution metallicity maps of eight large spiral galaxies observed as a part of the TYPHOON survey, producing more accurate predictive models of the metallicity distribution throughout these galaxies that can be applied to regions obscured by diffuse ionised gas (DIG) contamination where the metallicity cannot be directly observed. 

Despite these successes, a wealth of open questions remain. What are the main astrophysical sources responsible for these small-scale metallicity variations? Are they present in all galaxies, or only in large star-forming spirals in the local universe? Is the size of these small-scale variations set by the size ($R_e$; effective radius) of the galaxies? Do these small-scale metallicity variations also exist in passive galaxies, or dwarf galaxies? Are star-forming galaxies with strong spiral arms expected to be more or less well mixed than those without (i.e. grand design vs flocculent)? Are these small-scale variations stronger or weaker at cosmic noon, where galaxies are more clumpy and irregular and often show flat or inverted metallicity gradients? Finally, does the two-dimensional chemical structure of these galaxies match what is predicted by hydrodynamical simulations such as IllustrisTNG \citep{Pillepich+18} or FIRE \citep{Wetzel+23}? 

To answer these questions and others, it is important to extend these spatial statistical analyses to a larger sample of real and simulated galaxies, with a variety of sizes, morphologies, star formation rates, and redshifts. 
A large wealth of legacy IFS data exists \citep{Sanchez20} for which geostatistical analysis could potentially be applied. However, the application of geostatistical techniques to understand the ISM of galaxies has at present only been explored for high-resolution IFS (or IFS-like) data products. 

Resolution has been identified as the most important figure of merit in the design of an IFS survey, particularly when the goal is to capture detailed information about sharp structures that vary over small scales \citep{Mast+14}. 
Therefore, it is important to understand how a geostatistical analysis performs on data at a range of resolutions. This will provide useful information for the design of future surveys and numerical simulations if their aims include detailed analysis of the small-scale processes that determine ISM metallicity.

Similar studies have been done in the past exploring resolution limits in IFS surveys in the context of fitting metallicity gradients to galaxies. By rebinning IFS data and simulating observations with poor angular resolution, \citet{Yuan+13} found that metallicity gradients appear flatter when observed with low-resolution IFS (FWHM $\gtrsim 340$ pc for a system at z=1.49). These effects were further investigated and quantified by \citet{Acharyya+20} using simulated high-resolution IFS datacubes. They found that input metallicity gradients are generally recovered accurately when the FWHM of the instrumental PSF is smaller than $\sim 0.25 R_e$.

% Concerns have also been raised about 

% \citet{Baker+23} -- global properties of a galaxy may be important in determining the small-scale properties of a galaxy.
In this work, we investigate the impact of spatial resolution by following the methodology of \citet{Yuan+13} and \citet{Mast+14}. Specifically, we rebin high-resolution IFS data of three galaxies observed by the TYPHOON survey to increasingly coarse resolutions, in order to assess how the results of geostatistical analysis are sensitive to resolution. Further, we explore how the quality of model fits depends on the number of metallicity data points. Finally, we provide a first investigation into the quantitative effects of aperture bias by comparing models fit using unbinned spaxels associated with \Hii regions to models fit using integrated \Hii region data.

The structure of this paper is as follows. We discuss the data and how it is processed to produce low-resolution maps in Section \ref{sec:data}. In Section \ref{sec:methods}, we discuss our metallicity diagnostics, our methods for finding \Hii-dominated spaxels and \Hii regions, and the geostatistical model and fitting methods that we use. We present our results for NGC 5236 in detail in Section \ref{sec:results}. We show how the ability of our geostatistical pipeline to precisely and accurately recover details on the multiscale metallicity structure of these galaxies is affected by the number of available data points in Section \ref{ssec:vs_n_dp}, and by resolution in Section \ref{ssec:vs_res}. In Section \ref{ssec:vs_regions}, we explore the trade-off between using unbinned \Hii spaxels, which may be affected by aperture bias, and binned \Hii regions, which are fewer and necessarily at a lower spatial resolution than individual spaxels. To ensure our findings are not simply peculiarities of one galaxy, we repeat our analysis for two other spiral galaxies in Section \ref{sec:other_gals}. We discuss overall results in Section \ref{sec:discussion}, and summarise the main conclusions in Section \ref{sec:conclusions}, including general advice on when and how these geostatistical methods can be effectively applied to astronomical datacubes. Finally, we present our analysis based on alternate metallicity diagnostics in Appendix \ref{ap:other_diags}, a primer on some geostatistical techniques and concepts introduced earlier in this series in Appendix \ref{ap:extra_maths}, and a geostatistical explanation for how the variance of the metallicity measured in an IFS datacube depends on its resolution in Appendix \ref{ap:change_of_support}.

\section{Data} \label{sec:data}

We select NGC 5236 (M83) as our fiducial galaxy in this study, using high resolution spectroscopic data from the TYPHOON survey \citet{Poetrodjojo+19}. To ensure these results are valid for a range of galaxies, we also include data from two other galaxies in the TYPHOON survey, NGC 6744 and NGC 7793 (Seibert et al. in prep.). We summarise the properties of these galaxies in Table \ref{tab:gal_properties}.%, present detailed results for NGC 5236 in Section \ref{sec:results}, and present our main results for these other two galaxies in Section \ref{sec:other_gals}.

\begin{table*}
    \centering
    \begin{tabular}{llll}
        \hline
         Name & NGC 5236  & NGC 6744 & NGC 7793 \\
         \hline
         RA  &  13:39:55.96 &  19:09:46.10 &  23:57:49.83 \\
         DEC & -29:51:55.5  & -63:51:27.1  & -32:35:27.7 \\
         Distance (Mpc) & 4.9 & 11.6 & 3.6 \\
         $R_e$ (kpc) & 3.3 & 7.2 & 1.8\\
         $R_{25}$ (kpc) & 9.18 & 33.7 & 4.89\\
         Position angle & $54.0^\circ$& $13.7^\circ$& $94.4^\circ$\\
         Inclination    & $12.5^\circ$& $53.5^\circ$& $55.6^\circ$\\
         Morphology & Sc & Sbc & Scd\\
         Stellar Mass $(\log_{10} M/M_\odot)$ & 10.4 & 10.9  & 9.2\\
         Star formation rate $(M_\odot/$yr)& 4.0 & 4.0 & 0.25 \\
        \hline
    \end{tabular}
    \caption{Properties of the three galaxies explored in this paper. RA and Dec values are taken from the NASA/IPAC Extragalactic Database. Morphological classification and position angle data are taken from HyperLEDA. Inclinations are calculated based on the axis ratios reported by HyperLEDA, assuming a disc scale height of $q_z=0.2$ for all galaxies, following \citet{Battisti+17}. Distances, stellar mass estimates, and SFR estimates are taken from \citet{Leroy+19}. Half-light radii, also known as effective radii $(R_e)$ are taken from the $K_S$ band effective radii measured from the 2MASS large galaxy atlas \citep{2MASS}. B-band isophotal radii ($R_{25}$) are taken from \citet{deVaucouleurs91}. Both radii have been converted to physical units based on the adopted distance values.}
    \label{tab:gal_properties}
\end{table*}

Datacubes for all three of these galaxies were constructed using a Progressive Integral Step Method (PrISM) using the 2.5m du Pont telescope located at the Las Campanas Observatory in Chile (\citealt{Grasha+22}, Seibert et al. in prep.). A series of adjacent exposures were taken with a long slit of width $1.''65$, which were re-binned to psuedo-IFS datacubes with  a native angular resolution of $1.''65$. TYPHOON observations were only obtained when the seeing was smaller than the width of the long slit. We note that Las Campanas Observatory has a median seeing of $0.''6$ \citep{Persson+90}.

For NGC 5236 at a distance of $4.89$ Mpc \citep{Leroy+19}, an angular size of $1.''65$ corresponds to a physical pixel size of $39$ pc. Each datacube was rebinned to several additional resolutions by binning blocks of $f\times f$ adjacent spaxels together. We label each map according to its \textit{bin factor} $f$. In this study, bin factors of $1,2,4,6,8,10,12$ and $24$ are tested, corresponding to physical resolutions from $39$ pc to $940$ pc for NGC 5236. This selection of resolutions covers the range of physical resolutions from MUSE observations of local galaxies, to high resolution zoom-in cosmological simulations such as \textsc{auriga} ($125$ pc, \citealt{Grand+17}), to IFS observations of gravitationally lensed galaxies with the NIRISS instrument of JWST ($\sim 200$ pc, \citealt{Wang+22}), up to large IFS surveys of galaxies with coarse spatial resolution such as CALIFA ($\sim 800$ pc, \citealt{Esposina-Ponce+20}). Therefore, the range of spatial resolution considered allows us to explore for which, if any, of these kinds of data products and observational surveys a geostatistical analysis is feasible.

\begin{figure*}
    \centering
    \includegraphics[width=0.92\textwidth]{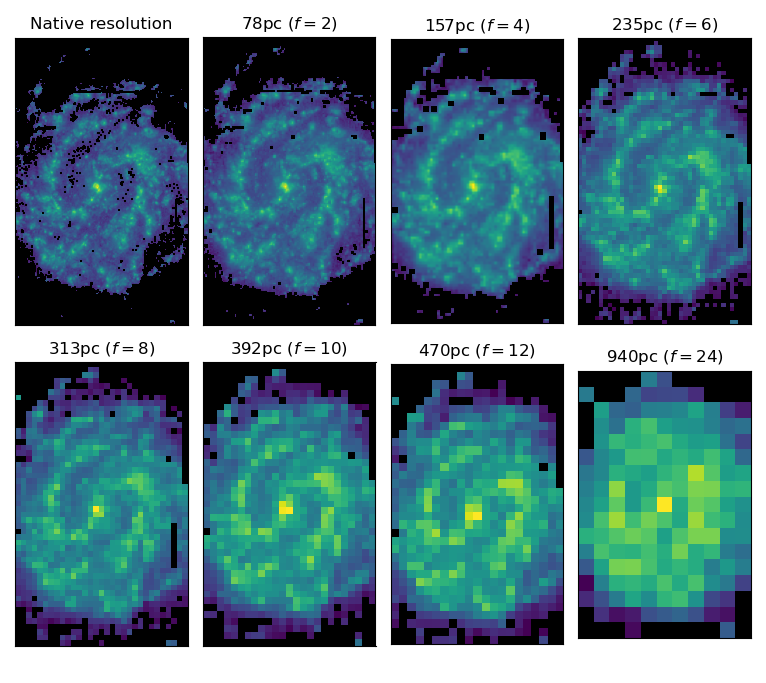}
    \caption{Surface brightness of H$\alpha$ for NGC 5236 at its native resolution (top left), and for spatially binned datacubes containing the total emission from $f \times f$ spaxels in each pixel. For the very high resolution maps ($f=1-2)$, individual \Hii regions may be resolved clearly. On larger scales ($f=4-6$), bright regions are blurred together, but small features are still visible. For intermediate resolution maps ($f=8-12$), small scale features become harder to see, but large-scale features such as spiral arms are still visible. With low-resolution IFU observations $(f=24)$, all features apart from a radial trend in H$\alpha$ brightness are lost.}
    \label{fig:postage_stamps}
\end{figure*}

To illustrate what features of the galaxy are visible, in Figure \ref{fig:postage_stamps} we show H$\alpha$ images for NGC 5236 at each resolution explored. At the native resolution, individual \Hii regions can be seen as ``bubbles" along the spiral arms of this galaxy, and regions of low H$\alpha$ emission in the inter-arm regions are visibly resolvable. As $f$ is increased, structural details of this galaxy are lost. At $f=4$ (with a spatial resolution of $\sim 150$ pc), individual \Hii regions begin to blur together, but variations can still be seen along spiral arms. These small-scale variations become difficult to see at $f=8$, but the azimuthally-varying spiral structure is still visible. At the coarsest resolution ($f=24$), no azimuthal structure is apparent, and a radial trend of decreasing H$\alpha$ brightness is all that can be seen.

\section{Methods} \label{sec:methods}

For each rebinned datacube, \textsc{lzifu} \citep{Ho+16} was used to find the intensities of 13 strong emission lines, including H$\alpha$, H$\beta$, [O\textsc{iii}]$\lambda5007$, [S\textsc{ii}]$\lambda\lambda6716,31$, and [N\textsc{ii}]$\lambda6583$.
The strength of all emission lines are corrected for dust attenuation assuming Case B recombination, fixing the Balmer decrement to be H$\alpha/$H$\beta=2.86$, using the extinction curve of \citet{ccm89}, assuming $R_V=3.1$ to match the Milky Way value.

\subsection{\Hii spaxels selection} \label{ssec:DIG}

Strong emission line ratio based metallicity diagnostics are only valid for \Hii regions -- regions recently ionised by hot, bright, young O/B stars. We ensure that all spaxels used in this analysis are predominately ionised by \Hii region emission at each resolution by imposing data quality cuts designed to isolate \Hii regions from diffuse ionised gas (DIG), a significant source of contamination for metallicity studies \citep[e.g.][\GeoGalsII]{Reynolds90, Poetrodjojo+19, Kumari+19}.

\begin{figure}
    \centering
    \includegraphics[width=0.49\textwidth]{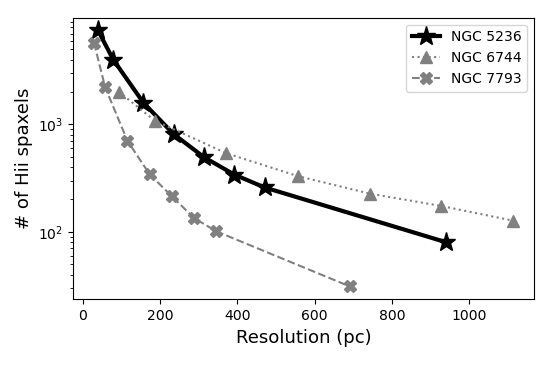}
    \caption{Number of \Hii spaxels as a function of resolution. For each galaxy, as the resolution of the data is decreased, the number of spaxels classified as \Hii dominated also decreases.}
    \label{fig:n_dp_per_res}
\end{figure}

We use two BPT diagram-based cuts \citep{BPT}, restricting our analysis to all spaxels that lie below both the \citet{Kewley+01} demarcation line on the [S\textsc{ii}]-based BPT diagram and the \citet{Kauffmann+03} demarcation line on the [N\textsc{ii}]-based BPT diagram. To ensure these cuts are robust, only spaxels with measurements of H$\alpha$, H$\beta$, [O\textsc{iii}]$\lambda5007$ [S\textsc{ii}]$\lambda\lambda6716,31$, and [N\textsc{ii}]$\lambda6583$ fluxes with S/N$ > 3$ are considered. Following \citet{Zhang+17}, we additionally impose a surface brightness cut, including only spaxels with a H$\alpha$ surface brightness greater than $10^{39}$ erg s$^{-1}$ kpc$^{-2}$. After these cuts, for NGC 5236, 7488 spaxels are classified as being predominately ionised by \Hii emission at native resolution. We hereafter refer to such spaxels as \Hii spaxels. As $f$ is increased to simulate coarser resolution observations, the number of \Hii spaxels recovered by this methodology decreases. We plot the number of \Hii spaxels available for analysis as a function of the resolution of the binned datacubes for all three galaxies analysed in this work in Figure  \ref{fig:n_dp_per_res}.

From observations of \Hii regions in the Milky Way, it has been established that the spatial extent of a \Hii region can range from tens of parsecs to kpc scales \citep{Cosens+18}, with typical sizes ranging from 80-800 pc \citep{Esposina-Ponce+20}. Therefore, the high resolution metallicity maps examined in this work contain \Hii spaxels that are significantly smaller than an individual \Hii region. 
This may introduce bias into measurements based on individual high-resolution spaxels, as emission line ratios are not constant throughout a \Hii region \citep[e.g.][]{Ferland+13}, and strong emission line based metallicity diagnostics are calibrated to work on integrated \Hii regions \citep[e.g.][]{Mannucci+21}.

To assess whether this bias is significant%\textbf{some of our high resolution data products}
, we additionally extract metallicities for integrated \Hii regions detected using the publically-available code \PHX\ \citep{Lugo-Aranda+22}. We choose this program over other \Hii region detection software because it does not require a license, it is actively maintained, and it was designed for optimal performance on high-resolution IFS data. Based on the H$\alpha$ intensity map and its uncertainty, positions and radii of hundreds of \Hii region candidates are extracted from each galaxy at native resolution. %\textbf{native resolution, OR A RANGE OF RESOLUTIONS, DEPENDING ON WHAT MICHELE THINKS IS INTERESTING}
The strengths of the emission lines for each candidate \Hii region are then computed by summing all (extinction-corrected) fluxes from spaxels contained within each region.

One limitation of \PHX\ for our application is that the same spaxel can be contained within two (or more) different overlapping extracted regions. In these cases, we assign spaxels to the \Hii region whose centre is closest to them. This step is necessary to prevent spurious correlations arising between neighbouring \Hii regions that are not physical, but merely statistical artefacts from double-counting.

We plot all \Hii region candidates on the two BPT diagrams, discarding any regions lying above the \citet{Kewley+01} demarcation line on the [S\textsc{ii}]-based BPT diagram and the \citet{Kauffmann+03} demarcation line on the [N\textsc{ii}]-based BPT diagram, leaving a final sample of \Hii regions for each galaxy. We note that this processing step decreases both the number of data points available for a metallicity model fit and the effective resolution of the data, but may be appropriate in some cases to mitigate against aperture bias. We explore how including this processing step affects a geostatistical model fitting procedure in Section \ref{ssec:vs_regions}.

\subsection{Metallicity diagnostics} \label{ssec:Z_diags}

To ensure that our results are robust, we investigate using four different metallicity diagnostics -- the \NSH\ diagnostic with the calibration of \citet{Dopita+16}, the Scal diagnostic of \citet{Pilyugin+Grebel16}, and the \ON\ and \RS\ diagnostics of \citet{Curti+20}. 
When using each diagnostic, we discard any \Hii spaxels (or regions) that have S/N$ < 3$ in any of the lines used. This final data quality cut excludes $<1\%$ of the data. 
%so it is not expected to introduce any substantial bias in the subsequent analysis. 
For all of the diagnostics we explore, we see similar trends for how the geostatistical methods tested are affected by resolution and data quality. We choose \RS $=\log( $[O \textsc{iii}]$\lambda5007/$H$\beta + $[S \textsc{ii}]$\lambda\lambda6717,6731$/H$\alpha $) as our fiducial diagnostic, presenting results from this diagnostic in the main text. This diagnostic is calibrated using direct electron temperature-based metallicity estimates of stacked spectra of SDSS galaxies with a 4th order polynomial fit for $Z$ as a function of \RS.
We refer the reader to Appendix \ref{ap:other_diags} for details on the other metallicity diagnostics investigated in this work, as well as for a presentation of the main results for NGC 5236 when these alternative metallicity diagnostics are used.

\subsection{Geostatistical model fitting}
\label{ssec:models}

In this Section, we introduce the hierarchical geostatistical model that we fit to our data, and the parameter determination procedure. The model is identical to the one introduced in \GeoGalsII. However, the fitting procedure has been updated and improved. For a summary on key background on  other tools from geostatistics that are used in this work and have beenintroduced in the earlier papers of this series, we refer the reader to Appendix \ref{ap:extra_maths}. 

For each data set that we test, we consider the observed metallicity $Z_{\rm obs}$ at each location $\vec{x}$ to be a combination of the true underlying metallicity at that point ($Z(\vec x)$), plus some uncorrelated, stationary, zero-mean observational noise $\epsilon(\vec x)$:

\begin{equation}
Z_{\rm obs}(\vec x) = Z(\vec x) + \epsilon(\vec x).
\end{equation}

We further separate $Z(\vec x)$ into two components: the process mean $\mu(\vec x)$, which captures the large-scale metallicity trends of the galaxy, plus a random component $\eta(\vec x)$ with zero-mean, which represents real, small-scale metallicity fluctuations around this mean:

\begin{equation}
\label{eq:breakdown_true_metallicity}
Z(\vec x) = \mu(\vec x) + \eta(\vec x).
\end{equation}

As in \GeoGalsI, \GeoGalsII, $\mu(\vec x)$ is modelled as being linearly dependant on the distance between $\vec x$ and the centre of the galaxy. To fully specify this model, two parameters must be fit: $\gradZ$, the radial metallicity gradient, and the characteristic metallicity $Z_{\text{char}}$ at a user-defined radius $r_{\text{char}}$:

\begin{equation}
    \mu(\vec{x}) = Z_{\text{char}} +  \gradZ \cdot \left( r(\vec{x}) - r_{\text{char}} \right).
    \label{eq:z_grad}
\end{equation}

In \GeoGalsI, \GeoGalsII, and in most studies when metallicity gradients are fit, $r_{\text{char}}$ is chosen to be zero. $Z_{\text{char}}$ therefore becomes the central metallicity of the galaxy. However, we found that such a choice of parameterisation led to strong anticorrelation between $Z_{\text{char}}$ and $\gradZ$, with $\rho < -0.9$ between these two parameters irrespective of the galaxy under investigation, the metallicity diagnostic used, or the resolution explored. Such a strong anticorrelation has been noted before in the literature, e.g. by \citet{Zaritsky+94}. 

To mitigate this anticorrelation, and in turn improve the efficiency of convergence using Monte Carlo methods, we parameterise the linear function by fitting the characteristic metallicity at a radius of $r_{\text{char}} = 0.4R_{25}$ (see \citealt{Zaritsky+94}). The metallicity at this radius has been observed to approximately match the integrated metallicity of face-on spiral galaxies using a variety of metallicity calibrations \citep[e.g][]{Moustakas+Kennicutt06, Rosales-Ortega11}. We found that such a parameterisation significantly lowered the degree of correlation between $Z_{\text{char}}$ and $\gradZ$, for all fits of our galaxy maps.

We parameterise the covariance of the small-scale, random metallicity fluctuations $\eta(\vec x)$ as an exponential function, following \GeoGalsII.
Such a functional form was chosen because in the small-scale limit, the structure of this field will mirror that of a passive scalar driven by Burger's turbulence \citep{Lee+Gammie21}, which is a good model for shock-induced structures \citep{Bec+Khanin07, Falceta-Goncalves+14}, and may be the most suitable paradigm of turbulence for describing the supernova shock-driven mixing of passive scalars around \Hii regions.
This function has two parameters: $\phi$, which sets the distance at which random metallicity fluctuations are correlated, and $\sigma^2$, the total variance about the process mean.
For two data points $\vec x, \vec y$ separated by a distance of $h$, we model the covariance between the small scale metallicity fluctuations to be:

\begin{equation}
    \text{Cov}(Z(\vec{x}), Z(\vec{y})) =\sigma^2  \exp \left( - \frac{h}{\phi}\right).
    \label{eq:matern_1/3}
\end{equation}

All four parameters of this hierarchical model ($Z_{\text{char}}, \gradZ, \phi$ and $\sigma^2$) are fit simultaneously using a maximum likelihood method, optimised using the Python package \textsc{emcee} \citep{emcee}. 
We refer the reader to Equation 6 of \GeoGalsII\ for the full form of the 4D likelihood function that is maximised. To help distinguish between values of $\sigma^2$ that are close to zero, and to ensure that the value of $\sigma^2$ was always positive, $\log \left( \sigma^2 \right)$ was fit instead of $\sigma^2$, with flat priors.

For every data set explored, we estimate posterior distributions for each parameter by running \textsc{emcee} for 480 steps using 72 walkers. Exploratory work showed that the autocorrelation time, $\tau_{AC}$, for each parameter is $\sim 40$ steps. Based on this, 480 steps was chosen such that the program could run for $10\tau_{AC}$ steps, after discarding $2\tau_{AC}$ steps for burn-in. This ensures there are effectively $\sim 720$ independent samples from the posterior distribution of each parameter, which is sufficient for estimating parameters with realistic uncertainties for this 4 dimensional problem. Each Monte Carlo chain was examined by hand to ensure that this burn-in length was long enough; for cases where it was not, an additional 40 steps were ran so that $3\tau_{AC}$ steps could be discarded.

For each run, walkers were initially started at a wide range of initial values of $\phi$ (ranging from $0.1$ to $1.6$ kpc) and $\sigma^2$, in order to ensure solutions were not confined to local minima, and a uniform prior was enforced on $\phi$ to keep it bounded between $50$pc and $5$kpc. The lower limit was chosen as it is similar to the size of a single pixel at native resolution for this dataset, and it is not possible to observe fluctuations that are smaller than a single pixel \citep{Cressie93}. For initial guesses of $Z_{\text{char}}$ and $\gradZ$ we use the results of a simple weighted-least squares (WLS) fit, with an uncertainty of $0.05$ dex on $Z_{\text{char}}$ and an uncertainty of $0.01$ dex/kpc on $\gradZ$, as central metallicities and metallicity gradients recovered for galaxies using the WLS method have been found to agree with those found with a geostatistical method to these levels (\GeoGalsII). After a test run, the chains of each Monte Carlo run were examined for convergence. We found that for some runs, all chains converged to low values of $\phi$ except for those that were started at values of $\phi$ that were much larger than the best-fit value. In these cases, we re-ran an MCMC fit using a set of initial values for $\phi$ ranging from $0.1$ kpc to $0.45$ kpc. In a few cases, the upper limit on the prior on $\phi$ was also lowered to $3$ kpc, to ensure all chains remained close to the best-fit value, as informed by previous runs. We note that adjusting parameter ranges based on information inferred from previous fits is a standard practice in a Bayesian framework, where the aim is to achieve quicker convergence on the global minimum while efficiently using the (limited) computational resources available.

\begin{figure*}
    \centering
    \includegraphics[width=0.95\textwidth] {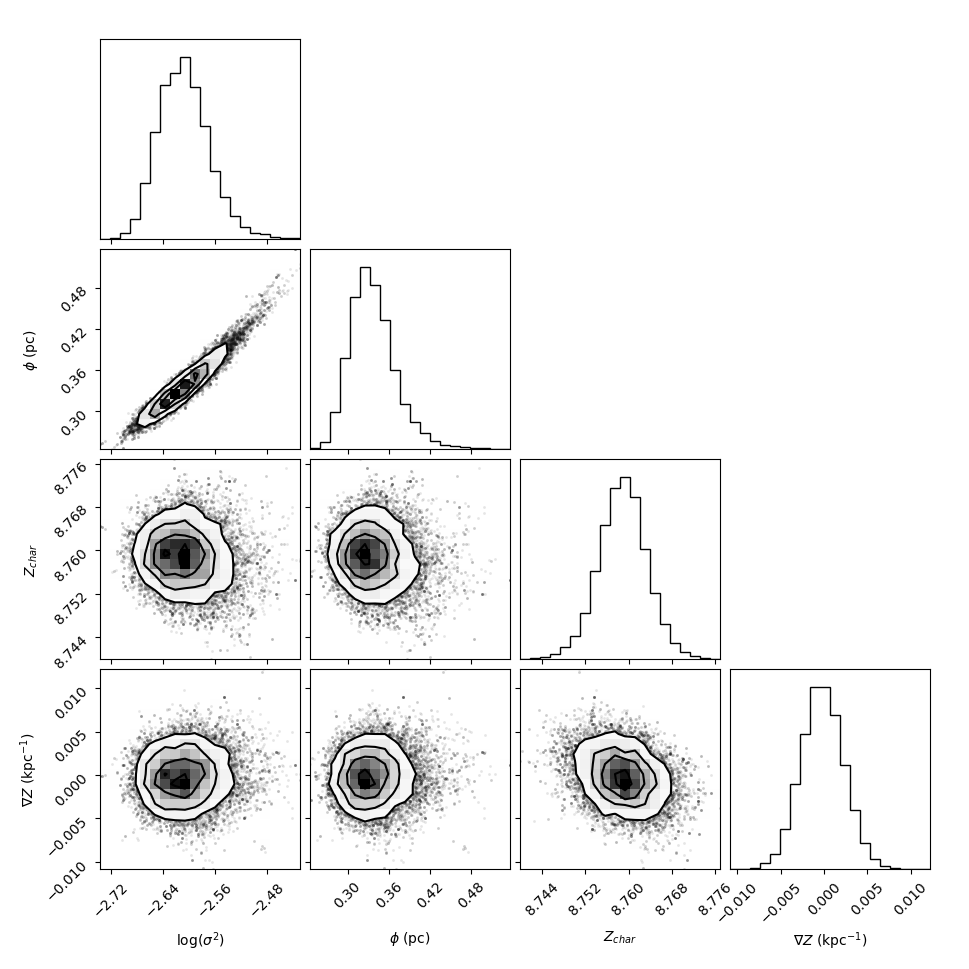}
    \caption{Corner plot, showing 1D and 2D marginalised posterior distributions for the four parameters of our geostatistical model, using all \Hii spaxels from our fiducial galaxy (NGC 5236), with our fiducial metallicity diagnostic (\RS), at the native resolution ($39$ parsec per spaxel). A strong correlation can be seen between $\phi$ and $\log(\sigma^2)$, the two parameters that describe small-scale metallicity fluctuations within the galaxy, as raising $\phi$ will naturally lower the variability of metallicity within the data. Such a correlation is seen in all \textsc{emcee} fits performed in this work.}
    \label{fig:corner}
\end{figure*}

To illustrate the quality of our fits, in Figure \ref{fig:corner} we show the approximate posterior generated by \textsc{emcee} as a corner plot for NGC 5236, using all \Hii spaxels, our fiducial metallicity diagnostic of \RS, and a datacube without any additional binning. From this plot, we see that the two small-scale parameters, $\phi$ and $\log(\sigma^2)$, are strongly correlated with each other ($\rho = 0.94$). As $\phi$ is increased, nearby data points are expected to become more and more correlated with each other, reducing the overall variability of the model. To match the amount of variation that is seen in the data set that is being fit, $\sigma^2$ must be increased to compensate for this effect. Therefore, models that fit the data and have a high value of $\phi$ must have a lower value of $\sigma^2$, and vice versa. Such behaviour is expected, and does not impact on the results presented in the following Sections.

\section{Results} \label{sec:results}

\begin{figure*}
    \centering
    \includegraphics{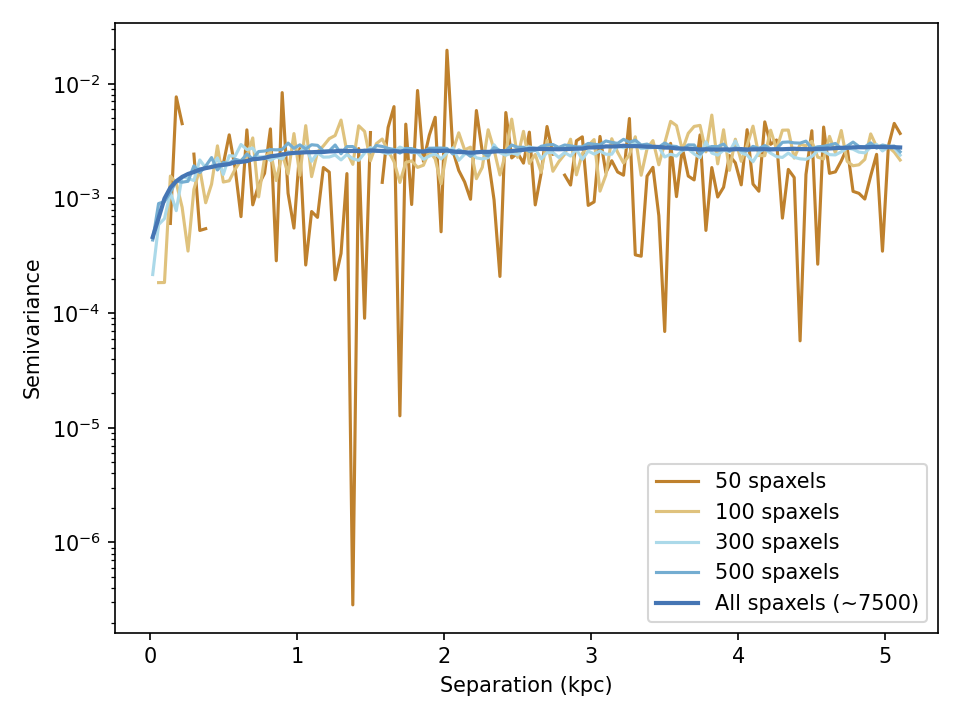}
    \caption{Recovered semivariograms for NGC 5236, when only a sub-sample of spaxels are used. In all cases, the semivariogram has the same shape -- however, when the number of data points is $\lesssim 100$, small number statistics lead to unreliable estimates of the variance between pairs of spaxels at each separation, increasing the noise in the semivariogram, making its interpretation less clear.}
    \label{fig:svg_vs_n_dp}
\end{figure*}

In this Section, we assess how the performance of the geostatistical methods explored for galactic metallicity modelling in \GeoGalsI\ and \GeoGalsII\ depends on three facets of the datacube used: the number of data points available, the resolution of the data, and whether the fit is performed on individual \Hii spaxels, or over integrated \Hii regions. However, separating out these effects is difficult. When spaxels are binned to make coarser datacubes, the number of data points available for statistical inference and model fitting decreases (see Figure \ref{fig:n_dp_per_res}). Furthermore, integrating \Hii spaxels into regions both decreases the number of available data points for model fitting, and the resolution of the available data points. 

With this in mind, in Section \ref{ssec:vs_n_dp}, we first assess how these methods are affected the number of data points with measured metallicity values available. We test this by randomly selecting subsamples of a limited number of \Hii spaxels from our flagship galaxy NGC 5236 at native resolution,  allowing us to change the number of available data points for analysis without introducing any additional spatial effects.

Then, in Section \ref{ssec:vs_res}, we report on how the model parameters recovered and the accuracy of metallicity predictions depend on the spatial resolution of the data by comparing results for our binned datacubes for NGC 5236 to the results found at native resolution. By comparing the results for the rebinned datacubes to the results on the native resolution data sets of Section \ref{ssec:vs_n_dp}, we can tell if there are any additional effects caused by having coarsely resolved data that would not be expected by merely changing the number of data points available.

Finally, in Section \ref{ssec:vs_regions}, we see if the results are the same for \Hii regions as they are for \Hii dominated spaxels, by comparing spaxels that pass the BPT-diagnostic cuts of \citet{Kewley+01} and \citet{Kauffmann+03} to regions found using \PHX\ that also pass these cuts (for further details, see Section \ref{ssec:DIG}). We interpret the results of this experiment in light of what we have learned from Sections \ref{ssec:vs_n_dp} and \ref{ssec:vs_res}, to see if binning the data into \Hii regions has any impact on the output of the model that cannot be explained by the other two effects investigated in this work.

\subsection{The effects of missing data} \label{ssec:vs_n_dp}

To test how geostatistical methods perform when the number of available \Hii spaxels is small, we generate random subsamples of 50, 100, 300, and 500 \Hii spaxels from NGC 5236. We compare the performance of our geostatistical methods on these subsampled datacubes to the full datacube, which has 7454 \Hii spaxels for which metallicities could be determined using \RS.

\begin{figure*}
    \centering
    \includegraphics[width=0.48\textwidth]{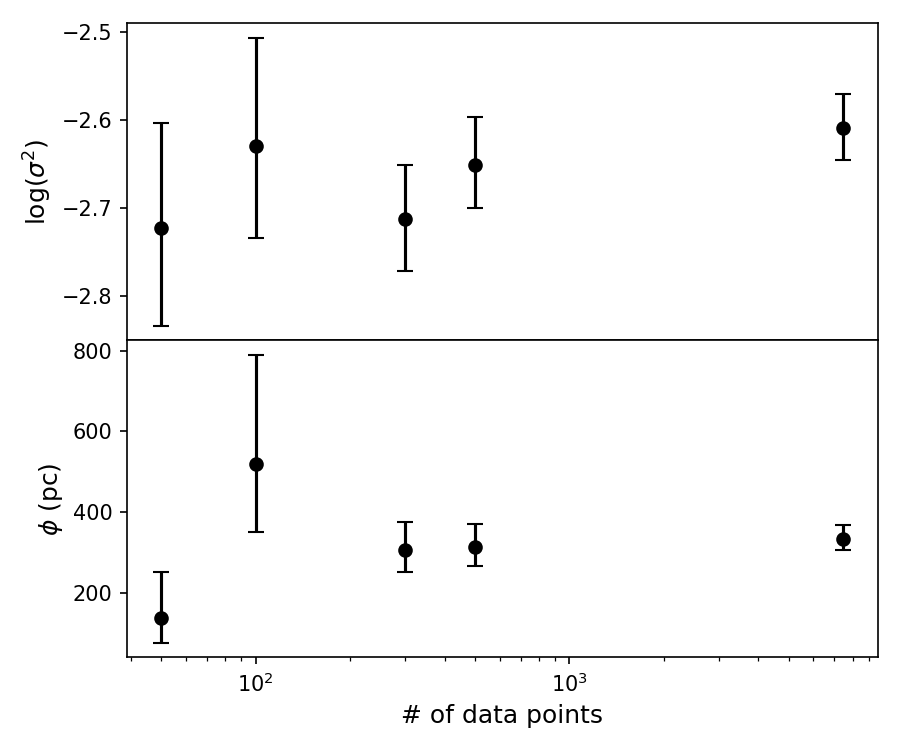}
    \includegraphics[width=0.48\textwidth]{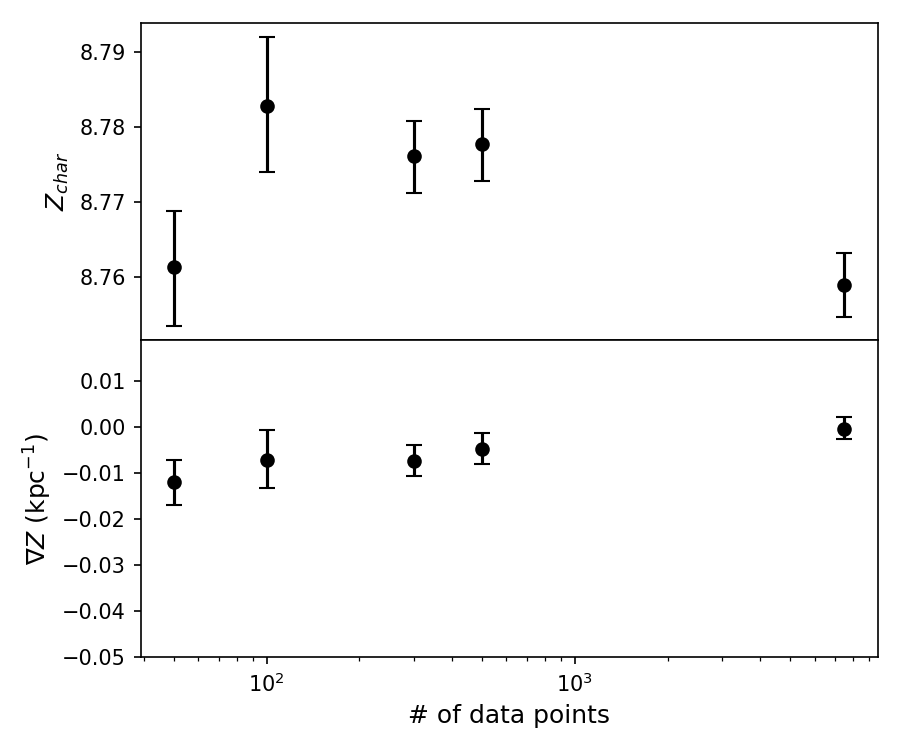}
    \caption{Best-fitting values, with uncertainty, of the metallicity correlation scale $\phi$, the (log) total variance $\sigma^2$, the characteristic metallicity $Z_{\text{char}}$, and the metallicity gradient $\gradZ$ for NGC 5236 as a function of the number of spaxels at native resolution used for the fit. Generally, we find that the best-fitting parameter values are consistent with the values recovered when all spaxels are used (far right data point), albeit with large scatter when the number of available data points is low.}
    \label{fig:param_convergence_vs_n_dp}
\end{figure*}

We create a semivariogram using each of the subsamples of the native resolution datacube in Figure \ref{fig:svg_vs_n_dp}, and plot them against the original semivariogram computed for this data. Each semivariogram shows how the variance in the metallicity between spaxels increases as the distance between spaxels is increased, after accounting for the large-scale metallicity gradient.\footnote{We provide a formal mathematical definition of the semivariogram, an introduction to this method, and its relation to some other tools used in astronomy to extract the covariance structure of random fields in Appendix \ref{ap:SVGs}.}
A bin size of $40$pc, chosen to match the resolution of the data and the minimum separation between spaxels, is used for all semivariograms.

We see that all semivariograms have the same general shape, indicating that the same small-scale structure is being detected in all five datasets. However, we also see that as the number of available data points decreases, the noise in the recovered semivariogram increases. When only $50$ data points are used, some separations have missing values for the semivariance, as there are not enough pairs of \Hii spaxels separated by approximately that distance for the variance between them to be estimated robustly. This issue could be partially overcome by creating a semivariogram with a coarser binning, allowing a more clear view of larger features, but doing so would result in less retention of information about the small-scale variability of the data. Since we are primarily investigating how many data points are necessary to recover these small-scale properties, we do not do this.

In Figure \ref{fig:param_convergence_vs_n_dp}, we compute estimates for all four parameters of our geostatistical model for each subsample of the native resolution data set, and investigate the convergence of these parameters. We see that even with as few as 50 data points, the properties of the large-scale variability ($\gradZ$ and $\Zchar$) are recovered accurately. In fact,  $\gradZ$ is consistent between all models, and $\Zchar$ varies by only $0.02$ dex. Similarly, all models converge to the same value for the amount of scatter around the mean trend ($\sigma^2$), although the uncertainty on this parameter increases as the number of available data points decreases. 

Recovered values on the metallicity correlation scale $\phi$ are consistent when $300$ or more data points are available for fitting. When only $100$ data points are used, the best-fit value of this parameter is larger, with larger uncertainties ($\phi = 518^{+271}_{-166}$ pc), but it is still consistent with the value found at native resolution ($\phi = 333^{+34}_{-28}$ pc) within the 1$\sigma$ error bars. For the model trained on only $50$ spatially resolved metallicity measurements, a smaller estimate of $\phi$ is obtained, with smaller error. By examining the posterior distribution of the Monte Carlo fit for this trial, we find that this is due to estimates clustering tightly to the lower limit of $\phi$ at $50$ pc, indicating that no evidence for spatial correlations is seen in this limited data set.

% {\color{red}{To rewrite when appendix comes in.}}
%When this analysis was repeated with alternative metallicity diagnostics, or using data taken for NGC 6744, large deviations of $Z_c$ and $\gradZ$ were not seen between the model fit to all spaxels and the models fit to subsampled datacubes. However, for NGC 7793, the trend of a flatter gradient appearing when more data was used to fit a geostatistical model was again recovered. In Section \ref{ssec:vs_res}, we see that such a flattening in the modelled metallicity gradient only happens for data with a very high resolution, and we provide a possible explanation for this effect in Section \ref{ssec:selection}.

Next, we evaluate how the goodness-of-fit of our geostatistical model is affected by having a limited number of data points. We judge the quality of our model fits over each subsampled datacube in terms of their predictive power. Following \GeoGalsII, we assess this using ten-fold cross validation. This method is a standard technique to estimate the ability of a model to generalise to unknown data points (\citealt{DiscoveringStats}, Chapter 7.7.2.), and is recommended as a reliable general method for model selection \citep{Hogg+Villar21}.
For each subsampled datacube, the \Hii spaxels are split into ten evenly sized groups. Data from all but one group are used to fit parameters for the geostatistical model described in Section \ref{ssec:models}, and generate point predictions for the metallicities of the \Hii spaxels in the final group via the process of universal kriging, outlined in Appendix \ref{ap:kriging}. This is repeated for all ten groups to produce a predicted metallicity for every spaxel. 

We summarise the results of this exercise by looking at the median absolute deviation (MAD) of the predicted metallicity of each spaxel to the metallicity observed for that spaxel, which we use as a metric for the model's performance. We then normalise this value to the variance of the data by dividing by one standard deviation, $\sigma$, to make this metric easier to interpret.\footnote{For reference, data from a Gaussian noise distribution around a known mean is expected to have MAD$/\sigma = \sqrt{\frac{2}{\pi}} \approx 0.8$.}
 
As a comparison point, we also repeat this cross-validation process for a metallicity gradient model with no small-scale component, fit for each set of data points using the standard weighted least squares (WLS) method, implemented in the Python package \textsc{statsmodels}. In the language of our model, this is equivalent to fixing $\eta(\vec{x}) = 0$. %Such a model does not capture any information on azimuthal variations. 

\begin{figure}
    \centering
    \includegraphics[width=0.5\textwidth]{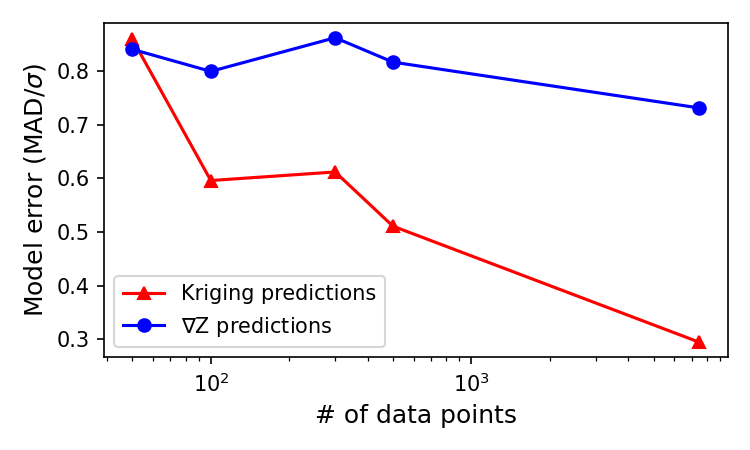}
    \caption{The normalised prediction error (median absolute deviation per standard deviation) of the predictions for metallicity computed using the cross-validation method for NGC 5236 using a simple linear gradient (blue line) and a geostatistical model (red line). As long as the number of data points $\gtrsim 100$, predictions are more accurate when a geostatistical model including small-scale metallicity variations was included.}
    \label{fig:prediction_error_vs_n_dp}
\end{figure}

We plot MAD/$\sigma$ for the linear gradient model and the geostatistical model as a function of the number of data points available in Figure \ref{fig:prediction_error_vs_n_dp}. For the linear model with no small-scale
component, we do not see a trend with the prediction accuracy model changing with the number of data points over the range we have chosen. This implies that $50$ data points is sufficient to construct a metallicity gradient model; and using more data points in the fit will not make this model any better. On the other hand, we see the accuracy of the geostatistical model scales with the number of data points available for training. With only $50$ spatially-resolved metallicities, the metallicity predictions of a geostatistical model have comparable accuracy to the predictions from a metallicity-gradient model. With $100-300$ data points, the geostatistical model begins to outperform the azimuthally-symmetric model that only captures large-scale metallicity variations within a galaxy. With a larger sample of data points, the improvement increases dramatically.

%{\color{red}{Point about how shortcomings/inaccuracies of the metallicity gradient only model impact astrophysical insights; e.g. SNe cosmology and H0.}}
This result is expected from geostatistical theory. Kriging works by taking advantage of the fact that nearby data points are expected to have similar properties (\citealt{Tobler1970}, Appendix \ref{ap:kriging}). When more data points are used, it is more likely that each data point lies within $\lesssim \phi$ parsecs of another spaxel with known metallicity, allowing the metallicity to be inferred using local galaxy properties as well as global ones. Furthermore, when more data points are used, the recovered values of the small-scale structure parameters $\phi$ and $\sigma^2$ are more accurate.

\subsection{The effects of spatial resolution}
\label{ssec:vs_res}

\begin{figure*}
    \centering
    \includegraphics{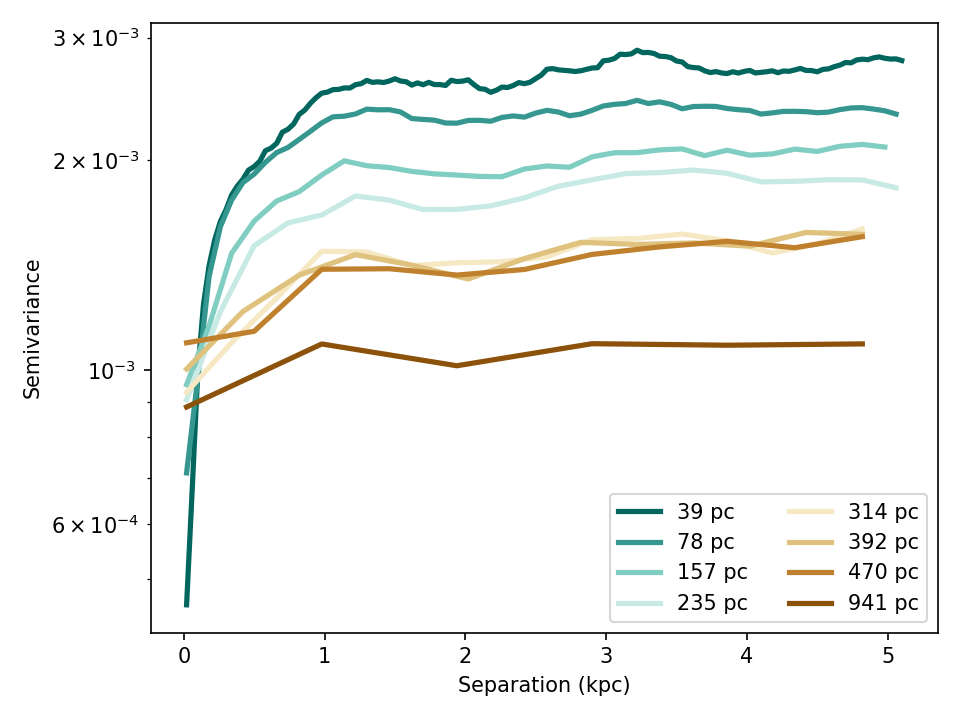}
    \caption{The semivariogram for the TYPHOON metallicity map of NGC 5236, binned to eight different spatial resolutions. As the resolution becomes coarser, the total variance in metallicity between spaxels decreases, as small scale inhomogeneities are binned together. When the spatial resolution is finer than $\sim 250$pc ($f \leq 6$), the shape of the semivariogram matches that of the native resolution map, indicating that the resolution is fine enough for small-scale metallicity fluctuations to be accurately resolved. For resolutions of $300-400$ pc per pixel, the general shape of the semivariogram can still be seen, and an estimate of the scale at which metallicities become locally uncorrelated can still be made. At resolutions coarser than this, it is difficult to infer the presence of small-scale fluctuations, and accurately describe their properties.}
    \label{fig:svg_vs_res}
\end{figure*}

To determine the resolution limit at which small-scale metallicity variations are visible, we construct a semivariogram for the small-scale metallicity variation using data convolved to each binning factor $f$. We plot these results in Figure \ref{fig:svg_vs_res}. 

The most striking trend that can be seen in this series of plots is that the height of the semivariograms at large separations decreases as $f$ is increased. This implies that the total amount of variance about a linear gradient is lower when the data is of a coarser quality. Qualitatively, this is to be expected. When the spatial resolution of IFU spectroscopy is coarser than the size of metallicity variations,  different features are averaged together over each individual spaxel, decreasing the variance around a galaxy-scale metallicity model. Quantitatively, we explore how this spatial averaging would impact the variance seen at different resolutions if the true small-scale metal variations follow an exponential correlation function with a correlation scale of $\phi=330$ pc, as is suggested by our highest-resolution data, in Appendix \ref{ap:change_of_support}. We find that the reduction of variance seen at different resolutions matches the reduction predicted with this model of spatially-correlated metallicity variations very tightly.

The separation at which the semivariance stops increasing (called the \textit{range} of the data in geostatistics literature; e.g. \citealt{Cressie90}) is not seen to increase with $f$ for all semivariograms computed with $f \leq 6$ (corresponding to physical resolutions finer than 250 pc per pixel, plotted as blue lines in Figure \ref{fig:svg_vs_res}). Visually, when $f \leq 6$, all semivariograms show the same shape, beginning to curve downwards at a separation of $\sim 0.4$ kpc, before flattening out completely by $\sim 1$kpc. For coarser data ($f > 6$, plotted as brown lines in Figure \ref{fig:svg_vs_res}), this shape is no longer clearly visible. At $f=8$, flattening is still seen at $\sim 1$kpc, but the shape of the semivariogram for sub-kpc data is no longer accurate. At $f>8$, we see the apparent range of the data increase with $f$, as trends associated with physical small-scale variations become washed out, and the size of the pixels becomes the most important scale for setting the observed structure of the metallicity map.

\begin{figure}
    \centering
    \includegraphics[width=0.5\textwidth]{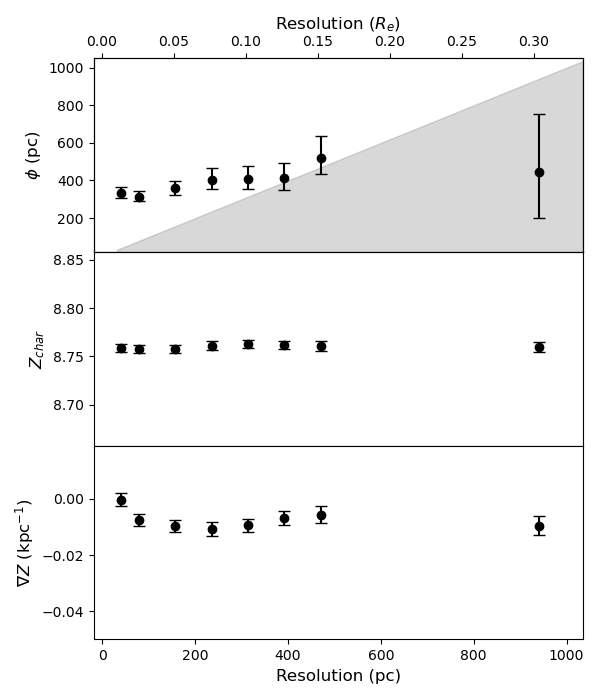}
    \caption{Best-fitting values, with uncertainty, of the metallicity correlation scale $\phi$, the characteristic metallicity $\Zchar$, and the metallicity gradient $\gradZ$ for NGC 5236 as a function of the spatial resolution of metallicity maps, reported in parsec (bottom axis) and effective radii ($R_e$, top axis). In the top panel, the grey shaded region shows the area of parameter space where the correlation scale $\phi$ is smaller than the size of a pixel. We find that kpc-scale resolved metallicity surveys are able to accurately recover large-scale trends of the metallicity profile ($Z_c,\gradZ$), while the correlation scale can only be accurately constrained by observations with a resolution $\lesssim 400$pc, or $0.12R_e$.}
    \label{fig:param_convergence_vs_res}
\end{figure}

To show this result in a more quantitative way, we present the best-fit value of $\phi$, the metallicity correlation scale, with uncertainty, as a function of the resolution of available data in Figure \ref{fig:param_convergence_vs_res}. Using the native resolution data, the most likely value of $\phi$ is $332^{ + 33 }_{ - 28 }$ pc. As $f$ is increased from $1$ to $10$, the best-fit value of $\phi$ increases slightly and has slightly larger uncertainties, but remains consistent with the value found for native resolution data. At $f=12$, the size of a single spaxel becomes $470$ pc -- larger than the value of $\phi$ recovered for the high-resolution maps. At this resolution, our model fitting procedure returns a value of $\phi= 527^{ + 135 }_{ - 90 }$ pc, inconsistent with the value of $\phi$ suggested by the higher-resolution maps. At our coarsest resolution, $\phi$ is inferred to be $443^{ + 308 }_{ - 244 }$ pc. While this value is consistent with the $\phi$ recovered at the finest resolutions, the uncertainties are very large, reflecting the fact that the value of $\phi$ cannot be well-constrained by this data, as a single pixel is expected to be much larger than the typical size of small-scale variations.

Figure \ref{fig:param_convergence_vs_res} also shows that the recovered values of $\Zchar$ and $\gradZ$ are relatively consistent throughout all data maps explored, excepting the results found for the datacube at native resolution, where a flatter metallicity gradient is preferred. We note that the fitted parameters for this model at native resolution are the same as those shown in Figure \ref{fig:param_convergence_vs_n_dp}. On its own, this result is not significant -- while the values of $\Zchar$ and $\gradZ$ recovered using our \textsc{emcee}-based parameter fitting pipeline do not agree with any other recovered values of $\Zchar$ and $\gradZ$ at other resolutions to $1\sigma$, we would not expect the $1\sigma$ credible intervals of eight different model fits to all overlap. However, we also find a similar result of the recovered value of $\gradZ$ being flatter when two out of our three alternative metallicity diagnostics are used (\NSH\ and Scal, with results inconclusive for \ON; see Appendix \ref{ap:other_diags}), and for the galaxy NGC 7793 when the size of a spaxel is $\lesssim 100$ pc (see Section \ref{sec:other_gals}). 
%and this does not indicate a failure of $Z_c$ and $\gradZ$ to be converged for data with a resolution coarser than $50$pc. At all other models, $Z_c$ and $\gradZ$ are consistent to within $1\sigma$, and models with larger recovered values of $Z_c$ (within uncertainty) have smaller values of $\gradZ$, consistent with the fact that these two parameters are highly anticorrelated.
%% new text
While formally significant, the size of this deviation is small -- smaller than the difference between metallicity gradients recovered for this galaxy when different metallicity diagnostics are used (see Figure 9 of \citealt{Poetrodjojo+19}).  

This result is unexpected from previous studies of the effects of resolution on recovering the metallicity gradient. In fact, previous research has shown that a flattening of the metallicity gradient is expected for poorly spatially-resolved data due to the effects of pixelisation and beam smearing \citep[e.g.][]{Yuan+13, Acharyya+20}. However, in this case, it is not a degradation of the data that causes an apparent flattening of the observed gradient, as a tendency for $Z$ to decrease as $r$ increases is still seen even in the native resolution datacube. 

\begin{figure*}
    \centering
    \includegraphics[width=0.75\textwidth]{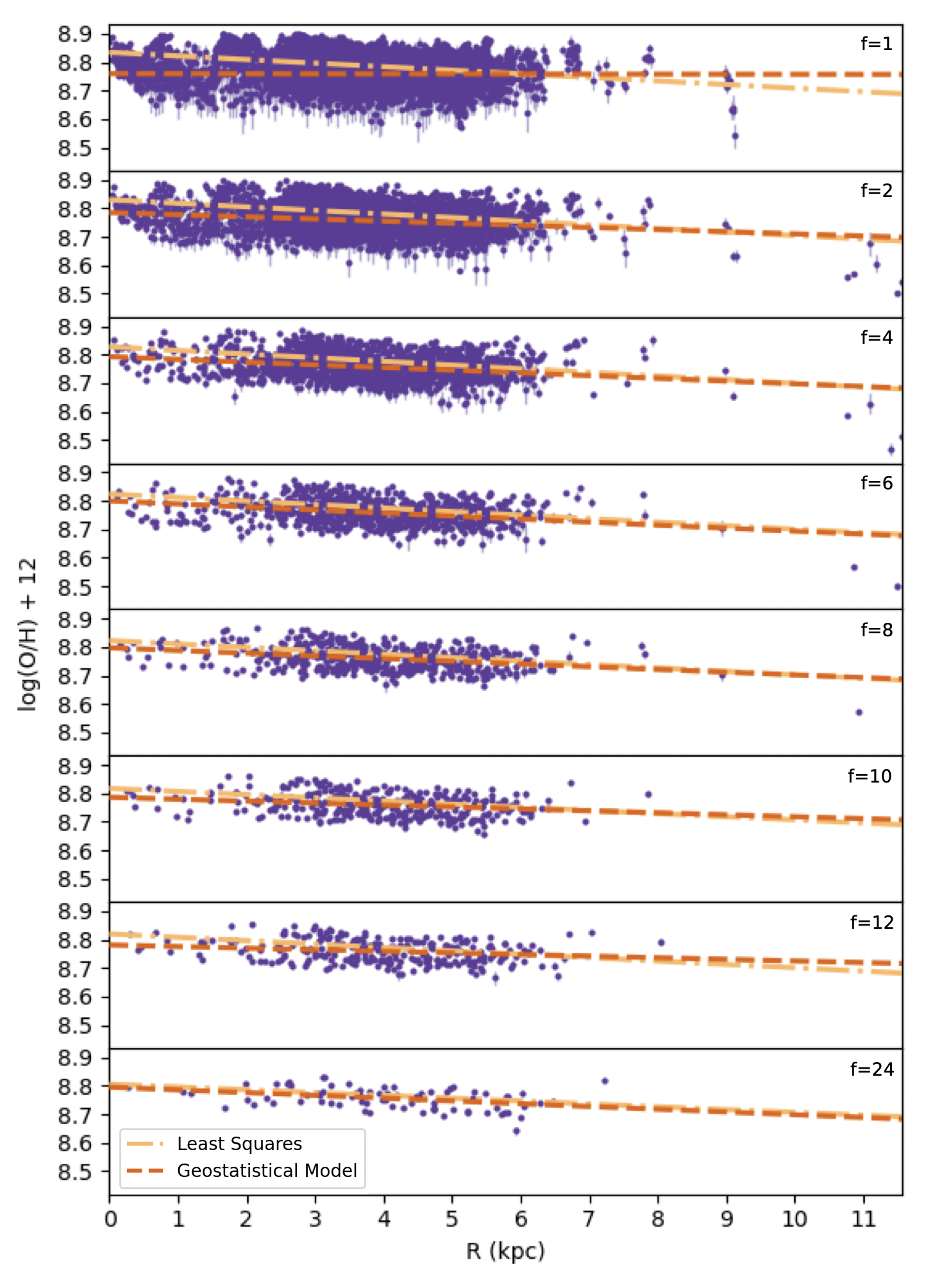}
    \caption{The radial metallicity profile of NGC 5236, computed using (i) our geostatistical methodology (red dashed lines) and (ii) the weighted least squares (WLS) algorithm (orange dashed lines) at different rebinned resolutions. When the WLS method is used, consistent values of the metallicity gradient are recovered for all binning factors $f$, with a slight tendency for the metallicity gradient to become flatter at the coarsest resolutions. For most of the resolutions tested, the radial metallicity profile recovered using our geostatistical approach agrees with the profile recovered using WLS. However, for the unbinned data ($f=1$), the geostatistical method prefers a flatter metallicity gradient.}
    \label{fig:Z_profiles_vs_res}
\end{figure*}

To test the validity of our statistical pipeline, we analysed the values of $\gradZ$ recovered using a simple WLS approach without incorporating any small-scale variation. We present our results in Figure \ref{fig:Z_profiles_vs_res}. At each resolution, we plot the measured metallicity of each \Hii spaxel, with error bars, at its distance from the galaxy centre. We overplot the best-fit linear relations describing the large-scale metallicity trends recovered using both our geostatistical approach, and using least-squares. For all of the binned data sets ($f > 1$), the profiles recovered using WLS and our geostatistical method show good agreement. However, when $f=1$, the geostatistical approach prefers to describe the data with a flatter metallicity gradient. This difference is significant, with $p < 5 \times 10^{-5}$ for the marginalised posterior value of $\gradZ$ being consistent with the least-squares value. Visually, at $f=1$, both lines appear to describe the data fairly well. While there is slightly more variance in the data points around the flat profile predicted by the geostatistical model, this is not unexpected given the different ways that these two models were optimised -- in the WLS approach, 
parameters for the line of best fit are chosen to minimise the scatter of all data points around the radial metallicity profile, 
whereas in our geostatistical model, data points are naturally expected to have some scatter around the large scale trend (for $f=1$, the most likely value for this scatter is $\sigma=0.05$ dex).

When the WLS method was used, the metallicity gradient recovered for the highest resolution map ($\gradZ = -0.0126$ dex/kpc) was 
within the range of metallicity gradients recovered for the other rebinned datacubes ($-0.013 \leq \gradZ \leq -0.010$ dex/kpc), with a slight trend of the metallicity gradient becoming flatter at coarser resolutions, as is expected from a coarser spatial sampling. However, for the \NSH\ and Scal diagnostics, an increase in the WLS-reported metallicity gradient was seen at the finest resolutions. We discuss a possible interpretation for this effect in Section~\ref{ssec:selection}.

%This indicates that the trend of metallicity decreasing as the distance from the galaxy center increases is still present in the finest resolution data for this galaxy. The information is not lost. Instead, our model describes this trend in a different way to the metallicity gradient model. Rather than modelling this variation as a simple, global, linear trend, at very fine resolutions and when a high number of spaxels with measured metallicity are available, our models prefer to fit a galaxy profile with a constant mean metallicity throughout, and describe the trend of metallicity decreasing with radius via a series of small-scale variations that together add up to produce this effect. 

\begin{figure}
    \centering
    \includegraphics[width=0.5\textwidth]{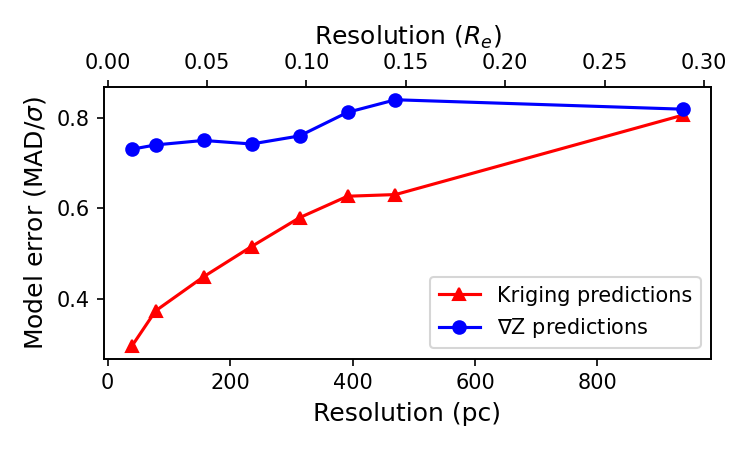}
    \caption{Normalised prediction error (median absolute deviation per standard deviation) for metallicity computed using the cross-validation method for NGC 5236 using a simple linear gradient (blue line) and a geostatistical model (red line). For all resolutions tested, predictions were more accurate when a geostatistical model including small-scale metallicity variations was included, with the greatest improvement seen at fine resolutions.}
    \label{fig:prediction_error_vs_res}
\end{figure}

In Figure \ref{fig:prediction_error_vs_res} we plot the normalised prediction error (MAD$/\sigma$, as calculated using ten-fold cross-validation) of the geostatistical model, and the linear model with no small-scale component, against the resolution of the datacubes. 
We find that when a linear gradient model is used, a similar prediction error value of MAD$/\sigma = 0.75-0.85$ is seen at all resolutions, with a slight trend of the predictive accuracy decreasing as the resolution is made coarser.
At all resolutions investigated, predictions from a geostatistical model were found to be more accurate than those from a metallicity-gradient model (although this improvement is only marginal for the coarsest resolution datacube with $f=24$). For resolutions finer than $\sim 330$ pc (the scale at which adjacent pixels are separated by a distance less than the metallicity correlation scale, $\phi$), the prediction accuracy of the geostatistical method improves rapidly as the resolution is improved. This highlights the power of this geostatistical modelling framework for being able to produce accurate predictive maps of the 2D metallicity distributions of galaxies, especially for high-resolution datacubes.

We note that while the geostatistical model fit to the highest resolution data has a value of $\gradZ$ that does not agree with the other models to $1\sigma$, this model is still able to predict the metallicity of \Hii regions with more accuracy than any other model. This indicates that the predictive accuracy of our geostatistical model is not severely impacted by the different values of $\gradZ$ recovered -- the model with a lower central metallicity and a flatter metallicity gradient is just as suitable for describing the observed data as a model with a stronger negative metallicity gradient. %By modelling the small-scale variations, and learning how metallicities are expected to be correlated between nearby data points, accurate predictions of the metallicities of \Hii regions throughout the galaxy can be made.

\subsubsection{Impact of spatial resolution on \Hii spaxel selection} \label{ssec:selection}

There are many possible reasons why the metallicity gradients recovered with the finest resolution spaxels exhibit small, but statistically significant deviations compared to metallicity gradients measured at simulated coarser spatial resolutions. 

First, it could be that the intrinsic radial trend in metallicity for these galaxies is not linear. High resolution metallicity maps have revealed breaks in a single metallicity profile that may not be visible in coarsely resolved data, potentially due to the influence of galactic bars \citep[e.g][]{Chen+23}. 
Flattening in metallicity profiles have been reported in the extended discs of spiral galaxies beyond $R_{25}$ \citep[e.g.][]{Rosales-Ortega11, Goddard+11}, including for this galaxy \citep{Bresolin+09}. However, the data used in this study is limited to a maximum galactocentric distance of $\sim R_{25}$, where flattening has not been found in previous studies of this galaxy's metallicity profile. Other studies have reported flatter metallicity profiles in the inner regions of galaxies \citep[e.g.][]{Belley+Roy92, Zinchenko+16}, but we do not see this in NGC 5236 at any resolution (see Figure \ref{fig:Z_profiles_vs_res}). 

The differences in inferred gradients could also be related to the selection of spaxels at each resolution. While we use a consistent \Hii spaxel selection method for all of our rebinned datacubes, the set of spaxels that are identified as being H\textsc{ii}-dominated is not the same between the emission line maps with different resolution. This effect is visible in Figure \ref{fig:Z_profiles_vs_res} -- at $f=2$, several spaxels are classified as H\textsc{ii}-dominated at radii of $\sim 11$ kpc that have no analogues in the $f=1$ map.

Additionally, (1) the \Hii region selection includes an empirically calibrated threshold from \citet{Kauffmann+03} based on Sloan Digital Sky Survey data that may not apply to apply to highly-resolved datacubes with resolutions finer than 100pc, and (2) the BPT diagnostic lines of \citet{Kewley+01} are computed to be the upper limits of the locations on the BPT diagrams that can be inhabited by \Hii regions -- however, no guarantees are made that spaxels that are not ionised by \Hii emission must lie outside of this region. Therefore, if a statistically appreciable number of spaxels outside \Hii regions are instead flagged by the DIG diagnostics as \Hii dominated for finely-resolved data, this could lead to a bias in the metallicity gradient computed for high-resolution data. 

To explore this hypothesis, we re-ran the analysis of Section \ref{ssec:vs_res}, ensuring that the same selection of spaxels were used at each resolution. To do this, we followed a procedure similar to the one outlined in \citet{Poetrodjojo+19}. First, using the native resolution data, for each metallicity diagnostic we define a mask containing only the spaxels that passed all data quality control cuts. When binning the data to coarser resolutions, only the spaxels that passed all data quality cuts at the finest resolutions were aggregated. For this test, we did not recompute line ratios for each rebinned spaxel with \textsc{LZIFU}. Instead, we simply summed the emission line fluxes of each spaxel within the mask contained within each binned spaxel, summing the errors in quadrature. 

Similar to \citet{Poetrodjojo+19}, and following \GeoGalsII, a harsher quality control cut for defining \Hii spaxels was used than the BPT diagram based methods. Based on the methods of \citet{Kaplan+16}, we compute the fraction of light in each spaxel that is coming from \Hii regions ($C_{\text{\Hii}}$) using the line ratio [S \textsc{ii}]/H$\alpha$. In more detail, using the native resolution maps, we assumed that the 100 spaxels with the brightest H$\alpha$ flux were entirely ionised by \Hii regions ($C_{\text{\Hii}}=1$), and the 100 spaxels with the faintest H$\alpha$ flux (that still have a S/N ratio greater than $3$ for both [S \textsc{ii}] and H$\alpha$) were entirely ionised by DIG ($C_{\text{\Hii}}=0$). We take the median values of [S \textsc{ii}]/H$\alpha$ of each of these populations to be the typical values of [S \textsc{ii}]/H$\alpha$ for \Hii regions and the DIG, respectively, and computed values of $C_{\text{\Hii}}$ for all spaxels by linearly interpolating where the measured value of [S \textsc{ii}]/H$\alpha$ falls between these two extremes. Spaxels with a value of $C_{\text{\Hii}} > 0.95$ are considered to be \Hii dominated and are included in our mask.\footnote{This threshold value of $0.95$ is (i) intrinsically arbitrary, and (ii) stricter than the threshold value of $0.9$ used in \citet{Poetrodjojo+19}. Such a conservative choice is consistent with the analysis presented in \GeoGalsII, where a 0.9 threshold was applied to $\log($[S \textsc{ii}]/H$\alpha)$, which for the TYPHOON emission line map of NGC 5236 is equivalent to a 0.95 threshold in [S \textsc{ii}]/H$\alpha$. 
%$C_{\text{\Hii}}$was erroneously calculated using measurements of $\log($[S \textsc{ii}]/H$\alpha)$ instead of [S \textsc{ii}]/H$\alpha$, with a threshold value of 0.9. Due to the monotonic nature of the log function, this is equivalent to using linear units with a different threshold value. For the TYPHOON emission line map of NGC 5236, this threshold value is 0.95088, so we choose to use a threshold value of 0.95 in this study to be consistent with our previous analysis.
} Additionally, any spaxels with S/N < 3 in H$\alpha$, H$\beta$, or any emission lines used in the metallicity diagnostic were discarded.

\begin{figure}
    \centering
    \includegraphics[width=0.5\textwidth]{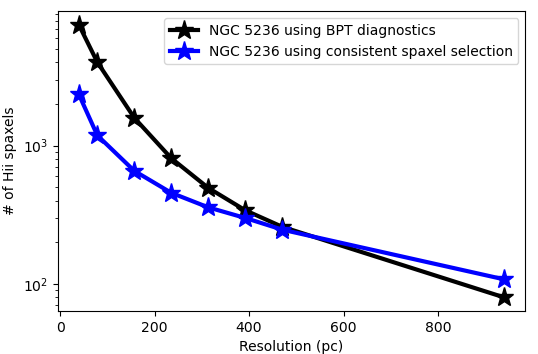}
    \caption{Number of data points available for NGC 5236 as a function of resolution, when \Hii spaxels are selected (i) using the BPT diagnostics of \citet{Kewley+01} and \citet{Kauffmann+03} together with the H$\alpha$ surface brightness cut of \citep{Zhang+17}; and (ii) by selecting a sample of \Hii spaxels at native resolution using the line ratio [S \textsc{ii}]/H$\alpha$, and rebinning this set of spaxels only to coarser resolutions. At the finest resolutions, the [S \textsc{ii}]/H$\alpha$-based cut removes many more spaxels than the BPT diagram based cuts. At coarser resolutions, however, more spaxels are retained, as binned spaxels that contain even one spaxel at native resolution that passes the data control cuts at native resolution are kept.}
    \label{fig:ndp_v_res_fixed_selection}
\end{figure}

Figure \ref{fig:ndp_v_res_fixed_selection} shows how the number of spaxels available for our flagship galaxy (NGC 5236) with our fiducial metallicity diagnostic (\RS) changes as a function of resolution with this new \Hii classification scheme. We compare this with the number of spaxels retained using the BPT diagrams and surface brightness cut as well. We see that, for the native resolution map, the [S \textsc{ii}]/H$\alpha$-based cut is much stricter than the BPT diagram-based cut: while $7488$ spaxels are classified as \Hii dominated at native resolution using the BPT diagram-based diagnostics, only $2351$ pass the [S \textsc{ii}]/H$\alpha$-based cut and have S/N > 3 for H$\alpha$, H$\beta$, [S \textsc{ii}]$\lambda\lambda6717,31$, and [O \textsc{iii}]$\lambda5007$. As the resolution is made coarser, the number of spaxels available for fitting decreases much less rapidly when the same set of spaxels are used at all resolutions. This may be due to an excess of spaxels being included at fine resolutions when the BPT-based spaxel selection criteria are used; or it could be due to spaxels being lost to DIG contamination at coarse resolutions.

\begin{figure}
    \centering
    \includegraphics[width=0.48\textwidth]{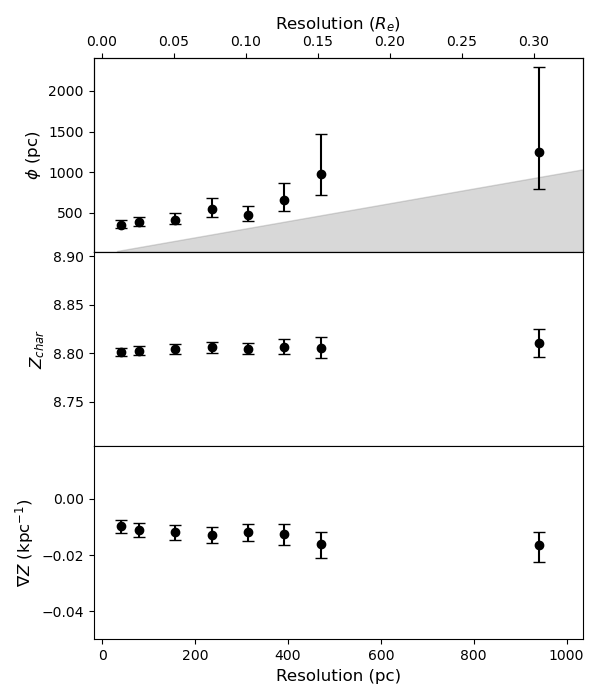}
    \caption{Parameters of our geostatistical model recovered with \textsc{emcee} after ensuring that the same set of spaxels are used at all resolutions. For this data, the value of $\gradZ$ recovered does not become significantly flatter as the resolution is changed.}
    \label{fig:param_convergence_vs_res_fixed_selection}
\end{figure}

Using a consistent selection of spaxels, but varying the resolution of the data, we then repeat our experiment of Section \ref{ssec:vs_res}, exploring how our set of recovered parameters vary with resolution. We plot our results in Figure \ref{fig:param_convergence_vs_res_fixed_selection}. We find that in this case, the metallicity gradient varies much less than was found for the case where \Hii spaxels were selected using a BPT-based approach.

A similar result was seen in the investigations of \citet{Poetrodjojo+19}. When DIG was excluded from the data at native resolution, metallicity gradients were found to be mostly constant with resolution. However, large variations in the metallicity gradient were found when different metallicity diagnostics were used, suggesting that systematic uncertainties in diagnostics are the most important limiting factor to achieve precise gradient measurements, rather than any effects resulting from having data with very fine spatial resolution. 

\subsection{The effects of aperture bias}
\label{ssec:vs_regions}

% \citet{Vogt+17} did a similar analysis. See consistent results with regions or spaxels.

% 

% Terlevich14: https://ui.adsabs.harvard.edu/abs/2014mysc.conf...67T/abstract "one cannot use global diagnostic diagrams when analysing “fractional” 3D data." -- they're talking about BPT tho, and this is just a conference result, and it doesn matter... 

In Sections \ref{ssec:vs_n_dp} and \ref{ssec:vs_res}, we assumed that all metallicities measured for \Hii spaxels are unbiased tracers of the underlying metallicities of the regions of the galaxy that they are drawn from. However, there are reasons to consider this may not be the case. Spherically-symmetric models of the interior structure of \Hii regions such as \textsc{Cloudy} \citep{Ferland+13} indicate that the ratio of emission lines within a \Hii region is expected to change as a function of the distance from the centre (see e.g. Figure 1 of \citealt{Mannucci+21}). For this reason, when the size of an IFS spaxel is smaller than the size of a \Hii region, the relative strengths of its emission lines are dependent not only on the metallicity within the spaxel, but also on the distance from the \Hii region's centre to the spaxel's centre, and on the relative sizes between the two. This may introduce an additional source of error in the metallicity estimated within each spaxel. 
By comparing the metallicities inferred from integrated DIG-free spectra to those that would be inferred from narrow slit spectroscopy, \citet{Mannucci+21} showed that in some circumstances aperture bias may have a stronger effect on recovered metallicities than DIG contamination. However, other studies with different methodologies have found the effects of aperture bias to be mild \citep[e.g.][]{Arellano-Cordova+22}.

\begin{figure*}
    \centering
    \includegraphics[width=0.8\textwidth]{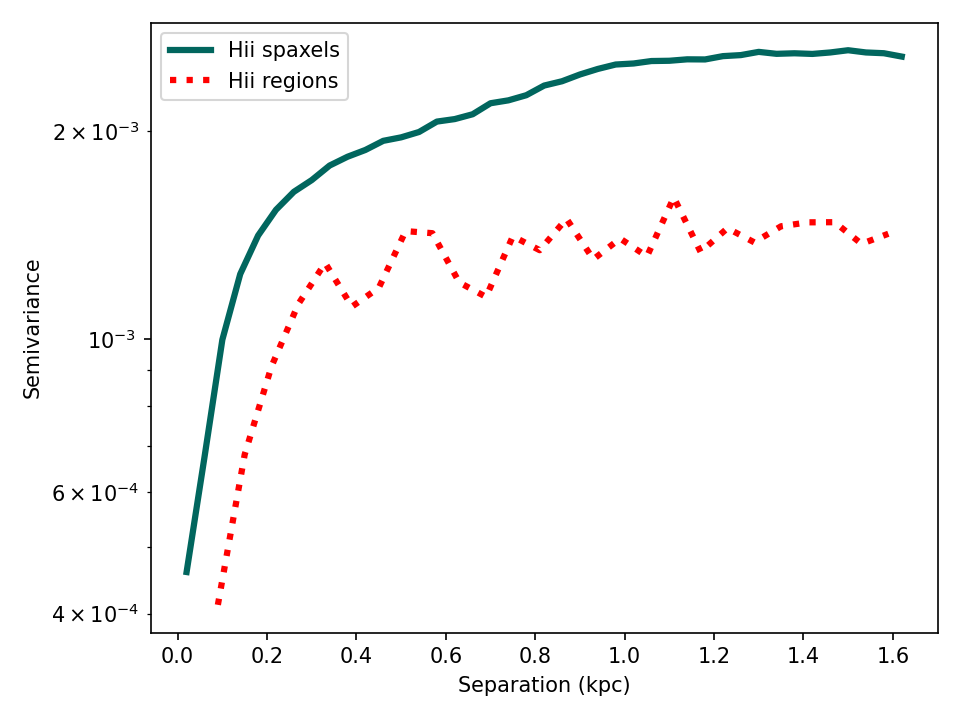}
    \caption{Comparing semivariograms calculated using \Hii dominated spaxels vs integrated \Hii regions found with \PHX. We see that individual spaxels carry more noise than \Hii regions, and show a small but noticeable increase in variance from $\sim 0.5-1$ kpc that is not seen in the region-based semivariogram. The additional fluctuations in the region-based semivariogram are likely caused by a low number of data points compared to the spaxel-based semivariogram.}
    \label{fig:svg_spaxels_vs_regions}
\end{figure*}

One step that can be added to an IFS data processing pipeline as a prophylactic measure against aperture bias is to first identify \Hii regions within a galaxy, and then integrate the emission from each spaxel within each region to get emission line ratios for each \Hii region. The line ratios for each region may then be converted into metallicities for each region. This ensures that no line ratio is artificially increased or decreased due to spatial sampling effects.

However, this method comes at some costs. Firstly, by binning multiple adjacent spaxels into single regions, the spatial resolution of the data is decreased, and the number of available data points that can be used to fit a 2D metallicity model is lowered. Secondly, many \Hii region-finding algorithms make assumptions on the size, brightness, and geometry of \Hii regions that may not be valid (see e.g. \citealt{Jin+22}), yield different results at different spatial resolutions \citep{Mast+14}, and usually depend on parameters that must be hand-tuned by the user, making the results of a \Hii region search difficult to replicate \citep{Lugo-Aranda+22}. 

In this context, a key question to ask is whether it is better to bin data into \Hii regions before beginning a geostatistical analysis, or to use higher-resolution, possibly biased data from individual \Hii spaxels.

In Figure \ref{fig:svg_spaxels_vs_regions}, we show semivariograms for the small scale metallicity variations within our fiducial galaxy (NGC 5236) using our fiducial metallicity diagnostic (\RS), comparing results found when all spaxels are used (blue line) to results using data binned to 485 \Hii regions (red line), found using \PHX\ that pass both BPT diagnostics (Section \ref{ssec:DIG}). We see that both semivariograms show the same shape, with the same flattening at $\sim 0.4$ kpc. Two other effects are also visible. First, the semivariogram produced using \Hii regions is more noisy. This is expected, because there are fewer data points available for the fitting -- see Section \ref{ssec:vs_n_dp} and Figure \ref{fig:svg_vs_n_dp}. Secondly, the height of the semivariogram in the large-scale limit computed using \Hii regions is shorter than the semivariogram computed using \Hii spaxels. This, too, is expected, because the size of a \Hii region is larger than a \Hii spaxel, and so averaging over a larger region reduces the amount of variance in metallicity (see Figure \ref{fig:svg_vs_res}, and further discussion in Appendix \ref{ap:change_of_support}). 

By comparing the height of the two semivariograms near zero separation, we see that the \Hii regions have far less uncorrelated variance than what is seen for \Hii spaxels. This observation has two possible interpretations. The first is that integrating over \Hii regions washes out small-scale metallicity variations that exist within an individual \Hii region. The second interpretation (which we consider more likely) is that metallicities estimated for integrated \Hii regions are more accurate than the metallicities estimated for individual \Hii spaxels. If this interpretation is correct, then by correcting for aperture bias, the error on the metallicity measurement of each spaxel is reduced. By comparing the height of these semivariograms near zero separation, the amount of variation reduction when this formulation is used could be estimated, potentially allowing the  effects of aperture bias to be quantified for a range of commonly-used metallicity diagnostics. We postpone such an analysis for a future study.

Overall, integrating spaxels into \Hii regions yields a metallicity map with fewer data points, coarser resolution, and more accurate metallicity measurements. In Table \ref{tab:params_spaxels_vs_regions}, we show what effect this integration step has on the best-fit values of $Z_c, \gradZ, \phi,$ and $\sigma^2$ recovered for this galaxy. We find that most of these parameters agree within their $68\%$ credible intervals.
The variance is found to be significantly lower for \Hii regions, which is an understandable consequence of the coarser resolution of the binned \Hii region data compared to the unbinned \Hii spaxel data (see Appendix \ref{ap:change_of_support}). A $3.6\sigma$ difference is seen between the best-fit values of $\Zchar$ for binned regions and individual spaxels, possibly pointing to a systematic error in metallicity measurements when this diagnostic is used on regions smaller than an individual \Hii region. However, the actual difference is very small, with the median values computed for regions and spaxels differing by only $0.02$ dex. This implies that the effect of such a systematic bias, if it exists, is small enough to be ignored, particularly given the larger (0.7 dex) uncertainties associated with strong emission line ratio based metallicity diagnostics \citep{Kewley+Ellison08}.

For the other two parameters, there are some slight differences within the bounds of uncertainty. For the \Hii region data, $\gradZ=-3.7\pm2.7 \times 10^{-3}$ dex kpc$^{-1}$. This is slightly more negative than the gradient found using native resolution \Hii spaxel data, but flatter than the gradients recovered for binned data, indicating that this data may be close to the limit of resolution at which the preferred models become those with no (or very weak) metallicity gradients, and variations within the galaxy is preferentially described through small-scale trends. Interestingly, the value of $\phi$ found for the \Hii region based data is lower than the value recovered when \Hii spaxels are used -- but the size of this effect is not strong enough to be significant.

\begin{table}
    \centering
    \begin{tabular}{l |c|c|}
        %\hline
         & \textbf{H\textsc{ii} spaxel fit} & \textbf{H\textsc{ii} region fit}  \\
         \hline
         $\Zchar$& $8.76 \pm 0.004$ & $8.78 \pm 0.004$\\
         $\gradZ$ ( 10$^{-3}$ dex kpc$^{-1}$) & $-0.2^{ + 2.5 }_{ - 2.4 }$& $-3.7 \pm 2.7$\\
         $\phi$ (pc)& $ 332 ^{+33}_{-28}$& $275 ^{+38}_{-32}$\\
         $\log_{10} \left( \sigma^2 \right) $ & $-2.61 \pm 0.04 $ & $-2.50 \pm 0.04 $ \\
         \hline
    \end{tabular}
    \caption{Best fit parameter results for the metallicity model for NGC 5236 when results are computed using data from each spaxel associated with a \Hii region, compared to results when using integrated emission line ratios from \Hii regions found using \PHX. In both cases, all spaxels/regions lie below the \citet{Kewley+01} and \citet{Kauffmann+03} demarcation lines in the S2- and N2-BPT diagrams. We find significant differences in the recovered values of two parameters for this galaxy with this diagnostic: the variance, which is understandable as \Hii regions are binned over a larger area than individual \Hii spaxels, naturally decreasing the variance in the metallicity between \Hii regions (see Appendix \ref{ap:change_of_support}), and the characteristic metallicity, $\Zchar$, possibly reflecting a small systematic bias when this diagnostic is used on spaxels smaller than an individual \Hii region.}
    \label{tab:params_spaxels_vs_regions}
\end{table}

Based on this analysis, we conclude that binning \Hii spaxels into \Hii regions does reduce the uncertainty of the metallicity estimated at each data point, and does not significantly impact the results of a geostatistical model fit. As long as the data remains at a resolution finer than the expected size of small-scale variations within a galaxy, and the number of data points remains high ($\gtrsim 100$), we advise binning \Hii spaxels into \Hii regions before fitting a geostatistical model. However, if this is not possible, there is no cause for concern, as a model fit using unbinned \Hii spaxels will likely yield similar results.

A similar conclusion was reached by \citet{Vogt+17}. They analysed the two-dimensional abundance distribution of HCG 91c captured by MUSE in two ways: (i) spaxel-by spaxel, and (ii) binning the data into \Hii regions using the open-source Python fitting code contained in \textsc{brutus}\footnote{\url{https://fpavogt.github.io/brutus/}}. They found that the metallicity gradient extracted was consistent when either spaxels or \Hii regions were used. Additionally, \citet{Vogt+17} noticed coherent, spatially-localised variations in metallicity on sub-kpc scales when both methods were used, representing an amplitude fluctuation of $0.2$ dex when computed from a spaxel-based analysis, and $0.15$ dex when the data was binned into \Hii regions. Such fluctuations agree with what we find in our analysis, with the spatially correlated variation remaining after spaxels were binned, albeit with a slightly lower variance. 

This result also shows that the fluctuations that we reveal with a semivariogram analysis are not caused by the varying ionisation conditions  within \Hii regions. Rather, they capture intrinsic variations in the metallicity of the ISM between \Hii regions. 

\section{Results for NGC 6744 and NGC 7793}
\label{sec:other_gals}

In Section \ref{ssec:vs_res}, we find that, based on the galaxy NGC 5236 and the \RS\ diagnostic, the limiting resolution at which our geostatistical pipeline can accurately recover the statistical properties of small scale metallicity fluctuations is $\sim 350$ pc. This value is comparable to the best-fit value of $\phi$, the scale of these fluctuations. Because of this, it is unclear as to whether this resolution limit depends on the size of the intrinsic fluctuations in NGC 5236, or is constant over a wider galaxy sample.
%depend on the size of the intrinsic fluctuations in NGC 5236?

We aim to make a qualitative assessment of the robustness of this resolution limit. For this, we expand our analysis to two other galaxies -- NGC 6744, an intermediate spiral galaxy at a distance of 11.6 Mpc, and NGC 7793, a flocculent spiral galaxy 3.6 Mpc away. Each of these galaxies was binned to increasingly coarser resolutions, using values of $f=1,2,4,6,8,10,$ and $12$. This corresponds to physical resolutions ranging from $93$ to $1113$ pc ($0.012-0.14R_e$) for NGC 6744, and $28$ to $346$ pc ($0.016-0.19R_e$) for NGC 7793. For NGC 7793, a binning with $f=24$ (corresponding to a physical resolution of $691$ pc) was also attempted, but only 31 spaxels satisfied our data cuts at this resolution, preventing a meaningful geostatistical model from being fit (c.f. Section~\ref{ssec:vs_n_dp}). 

\begin{figure}
    \centering
    \includegraphics[width=0.49\textwidth]{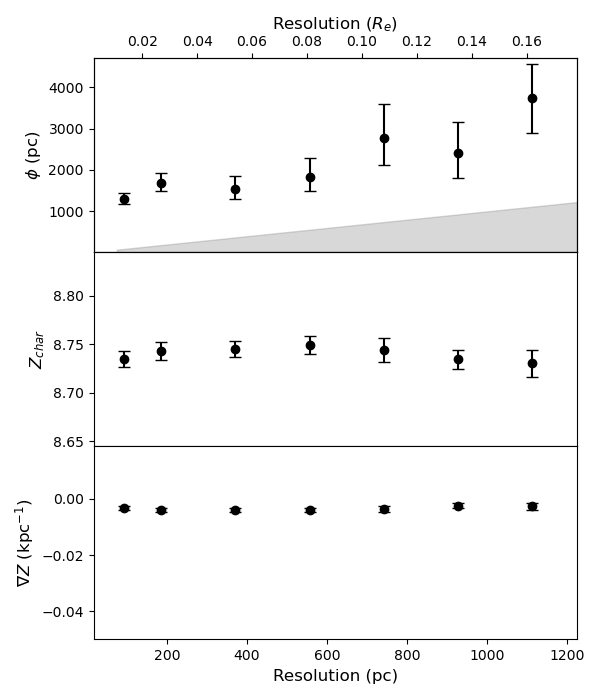}
    \caption{Best-fitting values, with uncertainty, of the metallicity correlation scale $\phi$, the characteristic metallicity $\Zchar$, and the metallicity gradient $\gradZ$ for NGC 6744 as a function of the spatial resolution of metallicity maps. The grey-shaded region in the top panel represents the region where $\phi$ is less than or equal to the spaxel size.}
    \label{fig:param_convergence_vs_res_N6744}
\end{figure}

\begin{figure}
    \centering
    \includegraphics[width=0.5\textwidth]{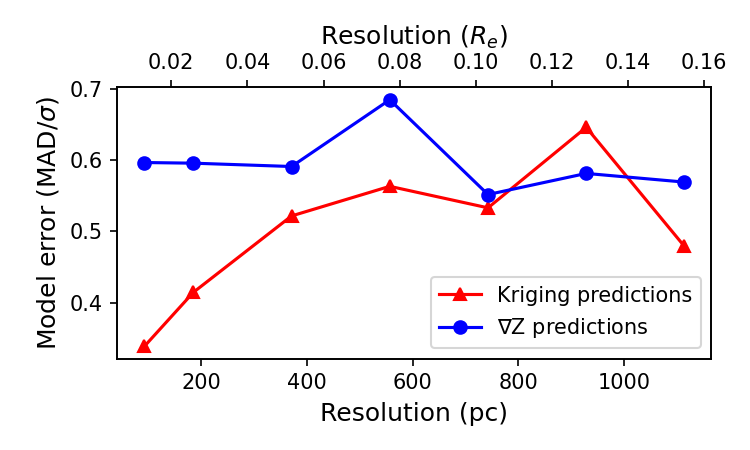}
    \caption{Normalised prediction error (median absolute deviation per standard deviation) of the predictions for metallicity computed using the cross-validation method for NGC 6744 using a simple linear gradient (blue line) and a geostatistical model (red line). When the small-scale structure can be resolved accurately, predictions from a geostatistical model are more accurate. However, at resolutions coarser than $\sim 750$ pc ($\sim 0.1 R_e$), kriging predictions are no more accurate than predictions from a linear model with no small scale component.}
    \label{fig:N6744_pred_error}
\end{figure}

\begin{figure}
    \centering
    \includegraphics[width=0.49\textwidth]{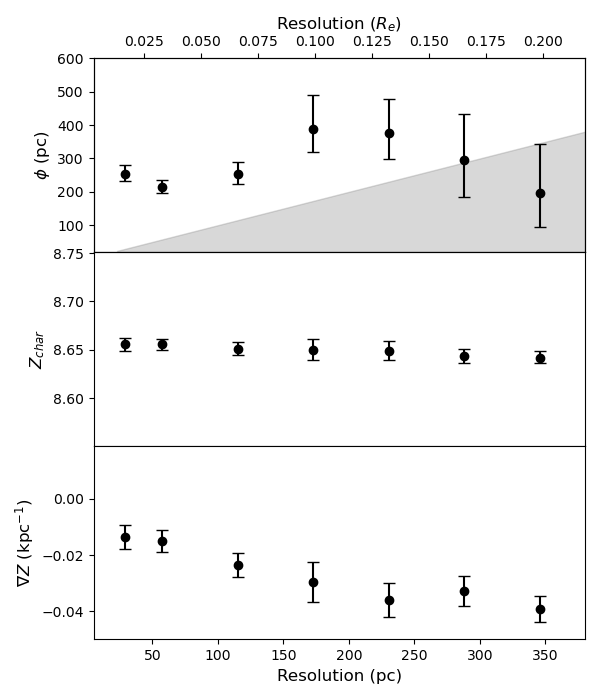}
    \caption{As in \ref{fig:param_convergence_vs_res_N6744}, but for NGC 7793.}
    \label{fig:param_convergence_vs_res_N7793}
\end{figure}

\begin{figure}
    \centering
    \includegraphics[width=0.5\textwidth]{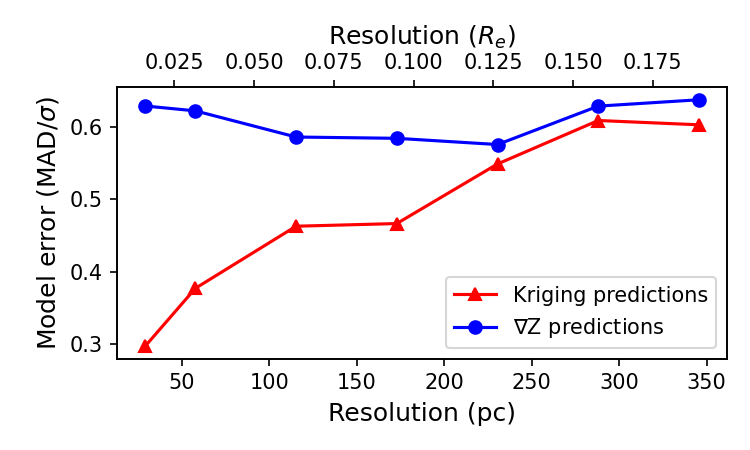}
    \caption{Normalised prediction error (median absolute deviation per standard deviation) of the predictions for metallicity computed using the cross-validation method for NGC 7793 using a simple linear gradient (blue line) and a geostatistical model (red line). When the small-scale structure can be resolved accurately, predictions from a geostatistical model are more accurate. However, at resolutions coarser than $\sim 200$ pc ($\sim 0.1 R_e$), kriging predictions are no more accurate than predictions from a linear model with no small scale component.}
    \label{fig:N7793_pred_error}
\end{figure}

At each resolution, a geostatistical model was fit to each galaxy using \textsc{emcee}. We show the median, 16th and 84th percentiles from our posterior distributions of $\phi$, $\Zchar$, and $\gradZ$ for NGC 6744 at each resolution in Figure \ref{fig:param_convergence_vs_res_N6744}, and for NGC 7793 in Figure \ref{fig:param_convergence_vs_res_N7793}, using \RS\ as our metallicity diagnostic. We also show the predictive accuracy of our geostatistical models using ten-fold cross-validation for NGC 6744 in Figure \ref{fig:N6744_pred_error} and for NGC 7793 in Figure \ref{fig:N7793_pred_error}. We compare the metallicities predicted for each spaxel via universal kriging to the measured metallicities of each spaxel, comparing it against the predictive accuracy a simple linear gradient model fitted via the weighted least squares (WLS) method.

For NGC 6744, at native resolution, the most likely value of $\phi$ is $1.30^{+0.15}_{-0.12}$ kpc -- much larger than the value of $\phi$ found for NGC 5236, even when NGC 5236 was binned to a resolution coarser than NGC 6744's resolution. This indicates that correlations between metallicities exist on much larger scales in this galaxy than in NGC 5236. Here, we see that the correlation scale $\phi$ is fairly stable to increasing values of $f$ until a resolution of $742$ pc per pixel is reached, comparable to half the value of $\phi$ found at native resolution. After this point, the best fit values of $\phi$ become significantly higher, and the size of their error bars increases dramatically. This in turn affects the predictive accuracy of this model, which we plot in Figure \ref{fig:N6744_pred_error}. After the point where the fit to $\phi$ stops being accurate, the geostatistical model becomes markedly worse, yet its predictive performance remains similar to a metallicity gradient model with no small-scale component.

For NGC 6744, the best fit values of the metallicity gradient and characteristic metallicity are stable for all resolutions tested, with all results consistent with each other within $1\sigma$ error bars. %A flattening of $\gradZ$ is seen at the finest two resolutions (corresponding to pixel sizes of $28$ pc and $56$ pc, respectively), indicating that, at these resolutions, the tendency of $Z$ to decrease as $r$ increases is being completely described by small-scale fluctuations.


For NGC 7793, at native resolution, the most likely value of $\phi$ is $ 254^{ + 24 }_{ - 21 }$ pc. Best fit values of $\phi$ are seen to deviate significantly from this value with large error bars when data with a resolution of $\gtrsim 170$ pc were used. This implies again that the result was stable to increases in resolution until a pixel becomes larger than $\sim \frac{1}{2} \phi$. 

We show how the predictive accuracy of our model is affected by resolution for NGC 7793 in Figure \ref{fig:N7793_pred_error}. At the two sharpest resolutions tested, the geostatistical model is able to predict metallicities at locations within the galaxy at a much greater accuracy than is possible with a simple linear gradient. At these resolutions, individual regions of enhanced star formation are well-resolved, and modelling small-scale deviations in metallicity captures much of the details of metallicity variability within this galaxy. At resolutions between $100$ and $200$ pc per spaxel, small-scale structure within this galaxy can still be marginally resolved and identified by our geostatistical pipeline, leading to models that, while not as successful at identifying local deviations from a large-scale metallicity trend, still exhibit a significant improvement in predictive power compared to a linear gradient model.
When this galaxy is observed with resolutions coarser than $\sim 200$ pc per spaxel, the size of a pixel becomes comparable to the typical size of a small-scale metallicity fluctuation, and any significant benefit in predictive power from fitting a geostatistical model is lost.

Overall, this analysis shows that the resolution requirements for a geostatistical analysis to be valuable does indeed depend on the spatial extent ($\phi$) of the small-scale variations within a galaxy. This makes coming up with general guidelines for the resolution requirements needed in order to study the internal metallicity distribution of galaxies using these geostatistical techniques difficult, especially when the typical size of small-scale fluctuations is not known in advance. However, we have noticed that the resolution at which geostatistical models begin to perform poorly often coincides with the resolution at which small-scale structure within a galaxy stops being apparent on a visual inspection. 

This leads us to suggest the following ``Golden Rule'' for when a geostatistical analysis is appropriate for IFS datacubes:  when small-scale deviations from a linear metallicity gradient appear to exist \emph{on visual inspection}, the size (in terms of both $\phi$ and $\sigma^2$) of these small-scale fluctuations can be captured accurately, and used to train predictive geostatistical models which can improve our understanding of small-scale metal mixing processes within galaxies. Our early results indicate that this is often achieved when the size of a resolution element is $\lesssim 0.1 R_e$. However, we caution the reader that this guideline is only based on observations of three large local star-forming spiral galaxies, and it is not yet known if this relation will hold true for dwarf galaxies, or galaxies with other morphologies.
At resolutions where small-scale fluctuations appear to exist but are not well-resolved (where $\phi$ is comparable to but smaller than the size of a pixel), geostatistical models offer some improvement in predictive power over a simple linear gradient model. Still, the best estimates of any parameters of the fit should be taken with some healthy scepticism, as poor resolution has been shown to bias $\phi$ to larger values, and lower $\sigma^2$ (see Appendix \ref{ap:change_of_support}).
Finally, when the data is so coarse that no structure aside from a radial trend can be seen, there will not be enough data to fit a geostatistical model that is more informative than a metallicity gradient would be on its own.

%\textit{to justify this ``Golden Rule" should I include pictures of the %galaxies N7793 and N6774 at different resolutions with Ha like Figure 1?}

\section{Discussion} \label{sec:discussion}

From the results presented in Section \ref{ssec:vs_res} and Section \ref{sec:other_gals}, we find that a resolution finer than $\sim \frac{1}{2} \phi $ is required to accurately recover the statistical properties of small-scale metallicity variations. From the literature (e.g. \citealt{Kreckel+19}, \GeoGalsI,  \GeoGalsII), small-scale fluctuations in ISM metallicity are seldom seen on scales larger than 1kpc (however, see \citealt{Williams+22})

For this reason, we do not consider it valuable to use this geostatistical methodology to analyse data products of $\sim 1$kpc IFS surveys, such as SAMI \citep{Croom+12}, MaNGA \citep{MANGA}, or CALIFA \citep{CALIFA}. This is unfortunate, as these surveys contain a large number of galaxies spanning multiple orders of magnitude in mass and star formation rate, which would allow investigations into how the properties of small-scale metallicity variations are set by a galaxy's global properties. 
However, there still exists a number of IFS surveys containing dozens of galaxies with spatial resolutions finer than $\sim 250$ pc. The combined catalogue AMUSING++ \citep{Lopez-Coba+20} contains IFS observations of 635 galaxies observed with spaxel sizes of $0''.2$ and a typical seeing of $1''.0$. Approximately $40\%$ of galaxies in this sample are close enough to be resolved with a seeing finer than $\sim 300$ pc, making this a statistically significant legacy dataset of IFS observations for which these geostatistical techniques could be applied. This dataset also contains galaxies of different morphologies. Analysing such a diverse population of galaxies using a geostatistical pipeline could provide insights into how the different drivers of turbulence affect gas mixing in a galaxy throughout its lifetime.

Similarly, geostatistical analysis could be applied to very high resolution cosmological simulations of galaxy formation. Specifically, applications would be ideal for zoom simulations such as FIRE \citep{Wetzel+23} and AURIGA \citep{Grand+17}, which simulate metallicity variations within galaxies on spatial scales as fine as $\sim 2$ pc. Such simulated datasets could readily be analysed with our geostatistical pipeline, in order to test the sub-grid physical prescriptions adopted in these simulations against observations, and thus both validate/falsify current choices and inform future theoretical models. 

Extending a geostatistical analysis to the population of galaxies at cosmic noon and beyond will be possible with the next generation of extremely large thirty meter class ground-based telescopes. The European Extremely Large Telescope (E-ELT; \citealt{ELT}) and the Thirty Meter Telescope (TMT; \citealt{TMT}) both include IFS instruments with resolutions of 4 milliarcseconds (HARMONI and IRIS, respectively) as part of their first generation suite of instruments. Under the $\Lambda$CDM cosmology of \citet{Planck18}, this angular resolution corresponds to datacubes with a physical resolution finer than 35 parsec per pixel at \emph{all redshifts}. Similarly, the IFS instrument GMTIFS planned for the Giant Magellan Telescope (GMT; \citealt{GMT}) will have an angular resolution of 6 milliarcseconds, which corresponds to a physical resolution of 52 parsec per pixel or finer at all redshifts. Based on the resolution analysis presented here, any one of these instruments will produce dataset suited for geostatistical analysis that will characterise the two-dimensional chemical structure of galaxies at high redshift to a high level of detail. Comparing such galaxies to local ones, it should be possible to determine whether the process of metal mixing is the same in these chaotic starbursting systems as they are in the Universe today. In turn, this will contribute to understanding whether the processes that govern stellar feedback are constant throughout cosmic time. With existing high redshift metallicity diagnostics \citep[e.g][]{Sanders+23, Laseter+23, Nakajima+23}, based on the wavelength coverage of these instruments, such analysis will be possible for galaxies up to $z \approx 4$. Progressing beyond this redshift limit is possible if new rest-frame UV metallicity diagnostics for high redshift galaxies are developed.

%This study also suggests that our geostatistical methods would not be useful on the output of large-scale, $\sim1$kpc resolution cosmological simulations of galaxy formation such as IllustrisTNG or EAGLE. 
%However, cosmological zoom simulations such as FIRE \citep{Wetzel+23} and AURIGA \citep{Grand+17} simulate metallicity variations within galaxies on spatial scales as fine as $\sim 2$ pc. Such simulated datasets could readily be analysed with our geostatistical pipeline, in order to test the sub-grid physical prescriptions supposed in these simulations against reality, and inform future theoretical models.

%discussion on puzzling results. 
Overall, our analysis highlighted the applicability and benefits of a geostatistical analysis. As with any novel diagnostic, the results obtained presents interesting aspects for further research. In particular, at the finest resolutions ($\lesssim 50$ pc per pixel), the best-fit geostatistical models generally tended to fit weaker metallicity gradients than was seen for data at coarser resolutions. This behaviour is not intuitive, as other studies have shown that binning data to coarser resolutions leads to flatter recovered gradients for galaxies \citep{Yuan+13, Mast+14, Acharyya+20}. However, these studies have focused on simulating galaxy observations with resolutions far coarser than the range in which this effect was seen to be important.

One interpretation of this effect is that mean metallicity gradients of galaxies are an emergent property that comes about from averaging over small-scale fluctuations. This would imply that the global metallicity properties of galaxies are set by their small-scale gas physics -- however, \citet{Baker+23} argue that kpc-scale features alone cannot be responsible for completely determining a galaxy's metallicity. Another interpretation is that the metallicity gradient of a galaxy is not an intrinsic property of the galaxy itself, and that higher resolution data reveals significant departures from a linear trend that require a different model to produce a good fit.

We find that a flexible geostatistical model accounting for small-scale metallicity fluctuations within the data produces very good predictions for the metallicity of unknown regions within a galaxy, especially with high resolution data. Obtaining accurate metallicity predictions at specific points of a galaxy is important not only for studies of galaxy evolution and ISM physics, but also for cosmology. Accurate estimates of the metallicity
at the site of many standard candles is required in order to calibrate their
period-luminosity relations \citep[e.g.][]{Ripepi+20, Bhardwaj+23}. Therefore, the predictive accuracy of metallicity models is directly linked to the precision of local Universe measurements of the Hubble constant \citep{Riess+22}.

The circularly-symmetric metallicity gradient model has been shown to be a good descriptor of the internal metallicity structure of local disc galaxies on the kpc-scale \citep[e.g.][]{Sanchez+14, Bresolin+Kennicutt15}. For these galaxies, such a model is well-motivated by theories of inside-out galaxy evolution. However, its applicability to merging systems, dwarf galaxies, and galaxies that display large asymmetries in their surface brightness profile is less well motivated. At high-redshift, galaxies are known to be more clumpy and irregular than local galaxies \citep{Conselice14}, suggesting that models with more flexibility than a linear gradient model may be important for understanding the baryon cycle at cosmic noon and earlier epochs. 
Understanding the data quality and resolution required to make a meaningful fit to a galaxy's metallicity map using a geostatistical model is an important first step to determining whether these models will be useful tools to analyse data from future IFS surveys of irregular systems and galaxies at high redshift.

In this work, the physical resolution of the data was defined to be the angular size of spaxels in psuedo-IFS data, times the distance to the galaxy. Distances to galaxies in the local universe can be highly uncertain. For example, we follow \citet{Leroy+19} in quoting the distance to NGC 6744 as $11.6$ Mpc, but other references have found the distance to this galaxy to be as close as $7.6\pm 1.7$ Mpc \citep{Sorce+14}. Such large discrepancies in the distances to these galaxies make the values of $\phi$ much more uncertain than the values that we report -- for example, adopting the distance from \citet{Sorce+14} would decrease the value of $\phi$ measured for NGC 6744 from $1.3$ kpc to $852$ pc. However, all results we have found on the relationship between the size of small-scale fluctuations and the resolution of the data would remain valid, as these two quantities scale with distance in a perfectly correlated way.

An alternative definition of the spatial resolution of an IFS datacube is based on the FWHM of the PSF for that data. For TYPHOON data, the PSF is several times smaller than the angular size of a spaxel. Because it is negligible at native resolution,  we do not simulate the effects of having a large PSF that extends over many pixels at any coarser resolutions, for consistency. When the PSF is much larger than the size of a spaxel in an IFS observation, metallicity gradients appear flatter \citep{Acharyya+20}, and small-scale features may be blurred or lost. Techniques such as forward modelling can be used to model the effects of poor seeing on large-scale galaxy properties and correct for them (Metha et al. in prep.), and help distinguish between physical small-scale features and small-scale features caused by a known PSF profile. Additionally, correlations expected between nearby spaxels caused by instrumental effects could be directly incorporated into a geostatistical model by changing the form of $\epsilon(\vec{x})$ to capture this correlated structure.

Another important effect that was not directly controlled for in the series of resolution experiments presented in this work is the effect of altering the signal to noise ratio, S/N, of the data. Metallicity maps with coarser resolution will have a greater S/N than those at finer resolution due to binning. Presumably, this would affect the performance of a model-fitting procedure. We note that for all resolutions tested, the same cut of S/N$>3$ on all lines used for metallicity diagnostics was enforced, ensuring that even at our finest resolution, no poor quality emission line fits are used to determine metallicities. We further note that geostatistical methods are suitable for distinguishing uncorrelated sources of noise, such as Poisson noise in a detector, from correlated signals of small-scale metallicity fluctuations. Thus, a geostatistical data analysis may be less impacted by having low S/N data than some other traditional statistical analysis techniques would be (see \GeoGalsI\ for further details). 

Finally, the inclination of a galaxy, $i$, is expected to make identification and classification of galaxy small-scale features more difficult, by effectively decreasing the resolution along one axis of a galaxy's plane. Understanding how the geostatistical models presented in this work are affected by inclination is important to determine which systems these techniques are best applied to, and for choosing targets for future high-resolution IFS surveys. Because a galaxy can only be seen from one inclination angle, and small-scale properties of the ISM are not uniform between galaxies, mock-IFS observations of simulated galaxies are required to investigate this effect. We leave investigations on this subject for future work.

\section{Summary and conclusions} \label{sec:conclusions}

In this series of papers, we introduce a novel framework for interpretation and modelling of astronomical datacubes based on spatial statistics techniques. In \GeoGalsI, geostatistics was introduced as a useful tool for extracting additional data from high-resolution metallicity maps, allowing local deviations from large-scale metallicity profiles to be detected, classified, and compared to theoretical models of metal mixing in a galaxy's ISM. In \GeoGalsII, we extended on this theme, showing how geostatistical models can be used to make accurate predictive maps of the metallicity structure of a galaxy in 2-dimensions, including within regions where DIG contamination prevents the metallicity from being directly measurable. Here, we investigate how the capability of geostatistical methods to observe, model, and predict small-scale features in the metallicity maps of galaxies are affected by data quality, to increase the sample size of (real and simulated) galaxies over which these methods can be applied, and aid the design of future surveys that seek to provide detailed descriptions of the ISM of a wide variety of galaxies. In particular, we focus on how the best-fit physical parameters and predictive power of a geostatistical model are affected by the resolution and completeness of an IFS-like observation. We summarise our main results below:

\begin{itemize}
    \item We found that $\sim 100$ data points with known metallicity are required to train a geostatistical model of metallicity variations within a galaxy's ISM. Below this value, noise due to small-number statistics makes computing a semivariogram very difficult, making it difficult to infer the presence of small-scale features. Furthermore, predictions from a geostatistical model perform no better than a linear gradient model. Increasing the number of data points available for a geostatistical analysis above this value increases the accuracy of kriging predictions, and decreases the uncertainty on the model's parameters.
    \item The resolution required to fit a geostatistical model that accurately captures small-scale features in the metallicity distributions of galaxies was found to depend on the size of those features. These features are typically of the order of $200-300$ pc across, but vary from galaxy to galaxy. As a general rule, we find that the geostatistical data pipeline presented in this work is able to capture any small-scale features that are apparent on visual inspection; and from the galaxies explored in this work, we find that this tends to be achieved when the resolution is better than $\sim 0.1 R_e$.
    \item For coarsely resolved data, the best fit value of $\phi$, the typical spatial extent of metallicity fluctuations, was found to depend on the resolution of the data. Therefore, best-fit values of $\phi$ should be trusted if and only if the size of a pixel is less than $\phi/ 2$. In these cases, predictions from a geostatistical model vastly outperformed a metallicity gradient model. 
    \item Even when the resolution of the data was not sufficient to accurately constrain $\phi$ to its true value, kriging predictions were still found to be much more accurate than predictions from a linear gradient alone for datasets with $\gtrsim 100$ data points, and pixel sizes $< \phi$ where small-scale structure is marginally resolvable by eye. When fewer data points are available, or the resolution is coarser, predictions from kriging offered only a mild improvement over predictions from a linear metallicity gradient.
    \item A tradeoff exists between using integrated \Hii regions and individual H\textsc{ii}-dominated spaxels for a spatially-resolved metallicity analysis. Integrating spaxels into \Hii regions decreases the number of data points available and the resolution of the data, but also reduces the uncertainty of the metallicity of each data point. Ultimately, we found that both methods yielded consistent results. Therefore, we
    recommend binning data points into \Hii regions before a spatial analysis only when the average sizes of \Hii regions stays below $\sim \phi/2$, and the number of data points stays above $\sim100$. 
\end{itemize}

Based on these results, there already exists a large legacy data set of local universe galaxy observations that have sufficient resolution and a sufficient number of data points for geostatistical models to be applied \citep{Lopez-Coba+20}. Key questions to answer include understanding how the metal mixing scale of galaxies varies with their size, morphology, star formation rate, and environment, to illuminate the interplay between small-scale gas physics and galaxy evolution. Comparisons to zoom-in galaxy simulations will allow detailed models of ISM mixing to be directly compared to observational data. Future observations with high-resolution IFS instruments, such as the first generation IFS instruments planned for E-ELT, GMT, and TMT, will allow geostatistical methods to be applied to galaxies at redshifts up to $z \sim 4$, revealing how turbulent mixing processes within galaxies have changed over cosmic time.

\section*{acknowledgements}
We would like to thank the anonymous referee, whose comments helped to improve the quality of this paper.
BM acknowledges support from an Australian Government Research Training Program (RTP) Scholarship. This research is supported in part by the Australian Research Council Centre of Excellence for All Sky Astrophysics in 3 Dimensions (ASTRO 3D), through project number CE170100013. BM would like to thank Prof. Tommaso Treu for his mentorship and advice on using MCMC methods and for hosting a wonderful extended research visit at UCLA. The authors extend their thanks to Alejandra Lugo-Aranda for the advice on using \PHX, and to Dr. Kathryn Grasha for the advice on using \HiiPhot and other helpful conversations. This research was conducted on Wurundjeri land.

\section*{data availability}
The TYPHOON data will be publicly released in Seibert et al. (in prep). Prior to this release, the TYPHOON data are available upon reasonable request to Andrew Battisti. Any further data products created for this work are available from the corresponding author (BM) upon reasonable request. A publically available Python package containing implementations of the geostatistical techniques explored in this series is currently in preparation (Metha et al. in prep.).

\bibliographystyle{mnras}
\bibliography{biblio} 

\newpage
\appendix

\section{Analysis with alternative metallicity diagnostics}
\label{ap:other_diags}

Estimating metallicities from strong emission line diagnostics has always been fraught with systematics. Different metallicity diagnostics can give metallicity estimates that differ by as much as 0.7 dex \citep{Kewley+Ellison08}. Furthermore, metallicity gradients measured in galaxies using different metallicity diagnostics can be different as well \citep{Poetrodjojo+21}.

For this reason, care must be taken to ensure that the findings of this study are not dependent on the specific metallicity diagnostic used. To ensure our results are robust, we test three
additional metallicity diagnostics, using three
different emission line ratios, calibrated in a variety of ways. The first of these is the \NSH\ diagnostic, together with the theoretical calibration of \citet{Dopita+16} devised using the plasma modelling code \textsc{Mappings 5.0} \citep{Sutherland+18}. This diagnostic uses three emission lines: [N\textsc{ii}$]\lambda$6583, H$\alpha$, and the doublet [S\textsc{ii}$]\lambda\lambda$6717,6731. This diagnostic has been shown to be largely insensitive to both dust extinction and variations in the ionisation parameter.

Secondly, we consider the S-calibration (Scal hereafter) of \citet{Pilyugin+Grebel16}. This empirical calibration was based on 313 reference \Hii regions with metallicities determined using the direct $T_e$-based method, and uses the emission lines H$\beta$, [N\textsc{ii}$]\lambda$6583, [O\textsc{iii}$]\lambda$5007, and the doublet  [S\textsc{ii}$]\lambda\lambda$6717,6731.

Finally, we examine how the results change when the \ON\ calibration of \citet{Curti+20} is used. This diagnostic was calibrated using exactly the same methods as our fiducial diagnostic, \RS. However, unlike the other diagnostics discussed in this work, \ON\ has no dependence on sulphur, instead depending only on H$\alpha$, H$\beta$, [O\textsc{iii}$]\lambda$5007 and [N\textsc{ii}$]\lambda$6853. Thus, it provides an independent test against effects that could artificially raise or lower the relative strength of sulphur lines, such as DIG contamination \citep{Haffner+09}.

\begin{figure}
    \centering
    \includegraphics[width=0.49\textwidth]{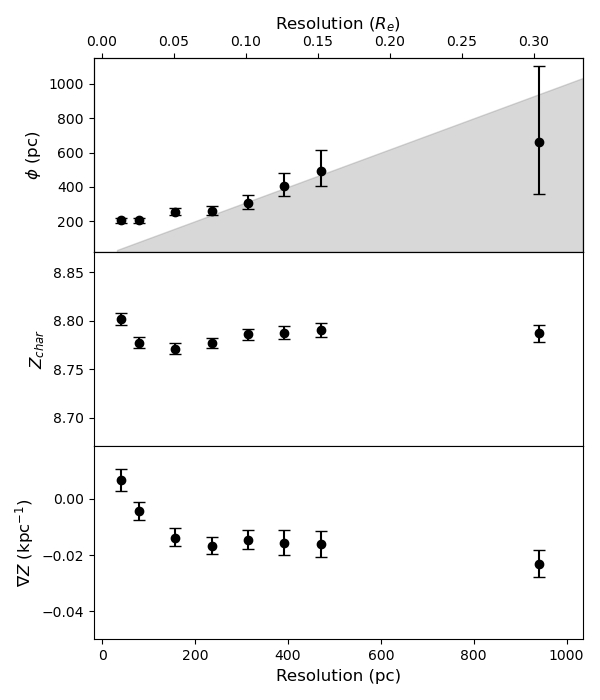}
    \caption{Medians, 16$^\text{th}$, and 84$^\text{th}$ percentiles of the posterior distributions of the parameters of the geostatistical model recovered for NGC 5236, when \NSH\ is chosen to be the metallicity diagnostic. At the finest resolutions, a model with a metallicity gradient close to zero is preferred. A value of $\phi \approx 200$ pc is preferred when this diagnostic is used; and this value can be accurately recovered until the pixel size becomes comparable to this value.}
    \label{fig:params_vs_res_N2S2Ha}
\end{figure}

We show how the fitted values of $\phi,$ $\Zchar$, and $\gradZ$ vary with resolution for NGC 5236 when the \NSH\ diagnostic is used in Figure \ref{fig:params_vs_res_N2S2Ha}. When this diagnostic is used, metallicity fluctuations are found to be correlated on a slightly smaller scale than was found when \RS\ was used, with $\phi=204 \pm 12 $ pc at native resolution. A consistent value is found when $f=2$, but for coarser resolutions with $f=4$ and $f=6$, this value increases slightly, to $\sim 250$ pc. At resolutions coarser than this, the recovered values of $\phi$ become increasingly high, as the size of a pixel becomes larger that the size of these small-scale fluctuations. We note a trend for the size of the fluctuations recovered to match the size of a pixel for four of the models fitted. This makes sense -- our binning process removes all signal of correlation below the size of a single pixel, but some small-scale correlations are still observable in the data, so the model assumes that the scale of the correlation and the scale of a single pixel are the same.

It is unclear to us why the recovered size of metallicity fluctuations are different when this diagnostic is used, as compared to \RS. Nonetheless, from this sequence of model fits, we see that to accurately recover the value of $\phi$, the size of a pixel must be smaller than $\sim \phi/2$ -- the same result that is seen for the two other galaxies discussed in Section \ref{sec:other_gals}. 

We also notice the same effect seen in Section \ref{ssec:vs_n_dp}, where the recovered value of $\gradZ$ is close to zero at our two finest resolutions. At coarser resolutions than this, the value of $\gradZ$ recovered is similar to, yet slightly steeper than, the metallicity gradient observed when our fiducial \RS\ metallicity diagnostic is used. Disparities in the metallicity gradient recovered when different metallicity diagnostics are used have been seen before for this galaxy \citep[e.g][]{Poetrodjojo+19}, and the size of the difference in our study ($\sim0.005$ dex kpc$^{-1}$) is consistent with such expectations of variability across diagnostic tools.

\begin{figure}
    \centering
    \includegraphics[width=0.5\textwidth]{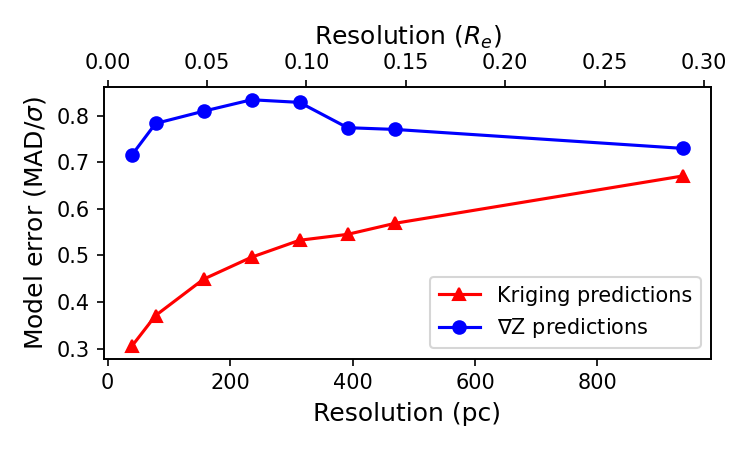}
    \caption{As in Figure \ref{fig:prediction_error_vs_res}, but using  \NSH\ as a metallicity diagnostic. We see that a geostatistical model fit still outperforms a linear gradient method at all resolutions tested.}
    \label{fig:N2S2Ha_pred_error}
\end{figure}

In Figure \ref{fig:N2S2Ha_pred_error}, we show the comparative predictive accuracy of our geostatistical model against a model that only contains a metallicity gradient at all resolutions tested. We see that our main results found when \RS\ was chosen to be our metallicity diagnostic also hold for \NSH: namely, (i) the geostatistical model outperforms a standard metallicity gradient model at all resolutions; (ii) the improvement in the fit is best seen for high-resolution data; and (iii) even when the recovered value of $\gradZ$ is close to zero for finely-resolved data, the predictions of the geostatistical model are still very accurate.

\begin{figure}
    \centering
    \includegraphics[width=0.49\textwidth]{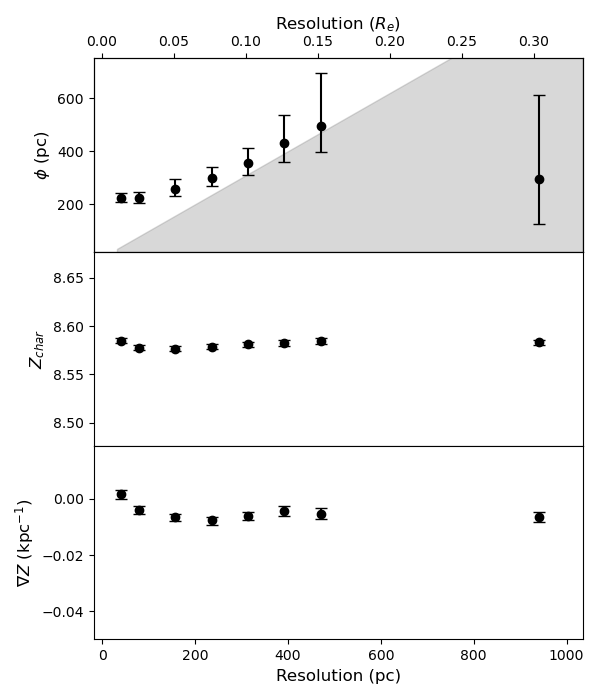}
    \caption{Medians, 16$^\text{th}$, and 84$^\text{th}$ percentiles of the posterior distributions of the parameters of the geostatistical model recovered for NGC 5236, when Scal is chosen to be the metallicity diagnostic. Values of $\phi$ recovered become unreliable when the pixel size approaches the size of the intrinsic fluctuations. At the finest resolution, the best-fit value of $\gradZ$ is close to zero, consistent with our other results.}
    \label{fig:param_convergence_vs_res_Scal}
\end{figure}

Our results when the Scal diagnostic is used are very similar to the results found when \NSH\ is used. This is to be expected, as these two metallicity diagnostics have been found to be highly correlated \citep{Groves+23}.
We show how the fitted values change as we change the resolution of our data in Figure \ref{fig:param_convergence_vs_res_Scal}. At native resolution, $\phi = 222^{ + 18 }_{ - 14 }$ pc, consistent with the value found when \NSH\ is used. Consistent values of $\phi$ are observed until the size of a pixel becomes larger than $200$pc. After this point, $\phi$ is consistently estimated to be equal to the size of one pixel, until $f=24$ where no small-scale structure can be found, and the best-fit value of $\phi$ is only recovered with very large uncertainties. As is found for the other two diagnostics tested, at the finest resolution, the geostatistical model prefers to fit a metallicity gradient that is close to zero.

\begin{figure}
    \centering
    \includegraphics[width=0.5\textwidth]{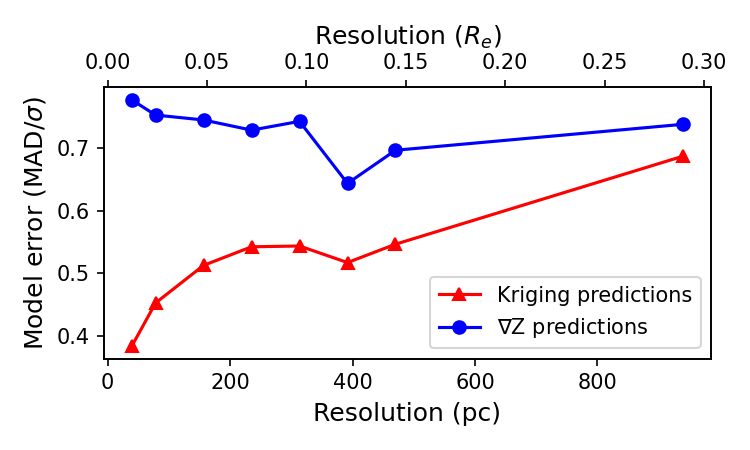}
    \caption{As in Figure \ref{fig:prediction_error_vs_res}, but using Scal as a metallicity diagnostic. Predictions from the geostatistical model always outperform those computed from the linear gradient model. The improvement is most significant at resolutions finer than $\sim 200$pc, comparable to the value of $\phi$ measured for this galaxy using this diagnostic.}
    \label{fig:Scal_pred_error}
\end{figure}

We plot the predictive accuracy of a geostatistical model for NGC 5236 when this diagnostic is used compared to a metallicity-gradient only model in Figure \ref{fig:Scal_pred_error}. We see the following results: (i) the geostatistical model always outperforms a linear fit; (ii) when the resolution is so coarse that $\phi$ cannot be accurately recovered, the improvement over a linear gradient fit does not improve as the resolution of the data improves; (iii) when the data is of a fine enough resolution for $\phi$ to be recovered accurately, the predictive accuracy of the model continues to increase as the resolution is made finer, all the way down to native resolution; and (iv) having a recovered gradient close to zero at the finest resolution does not affect the predictive performance of this model.

%\textbf{If the rest of our diagnostics are in agreement that $\phi \sim 200$ pc, then why would we make our fiducial diagnostic one that does not agree with this? With this in mind, do you think Scal or something should be promoted to be our new fiducial diagnostic?}

\begin{figure}
    \centering
    \includegraphics[width=0.49\textwidth]{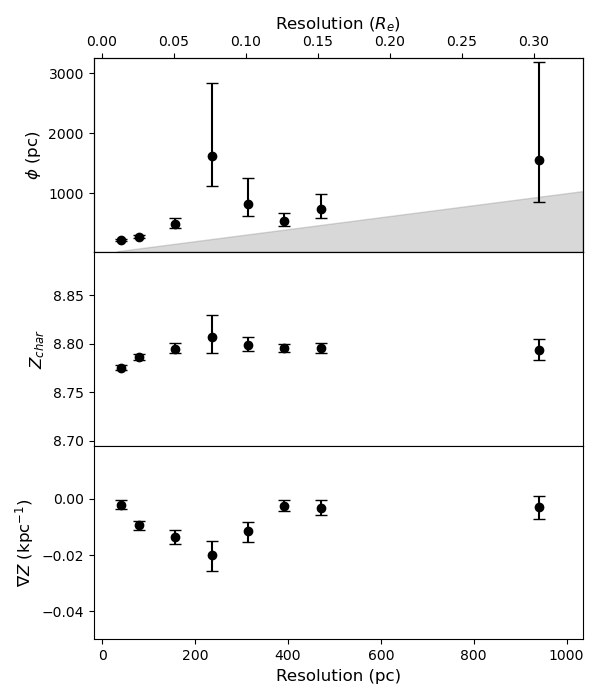}
    \caption{Medians, 16$^\text{th}$, and 84$^\text{th}$ percentiles of the posterior distributions of the parameters of the geostatistical model recovered for NGC 5236, when \ON\ is chosen to be the metallicity diagnostic. When $f=6$, small-scale structures stop becoming apparent in the data, as evidenced by the large value of $\phi$. Around this point, the best-fit value of $\gradZ$ decreases and $\Zchar$ increases. For smaller and larger resolutions, the trend of $\phi$ increasing with resolution is similar to what is seen in other diagnostics.}
    \label{fig:param_convergence_vs_res_O3N2}
\end{figure}

\begin{figure}
    \centering
    \includegraphics[width=0.5\textwidth]{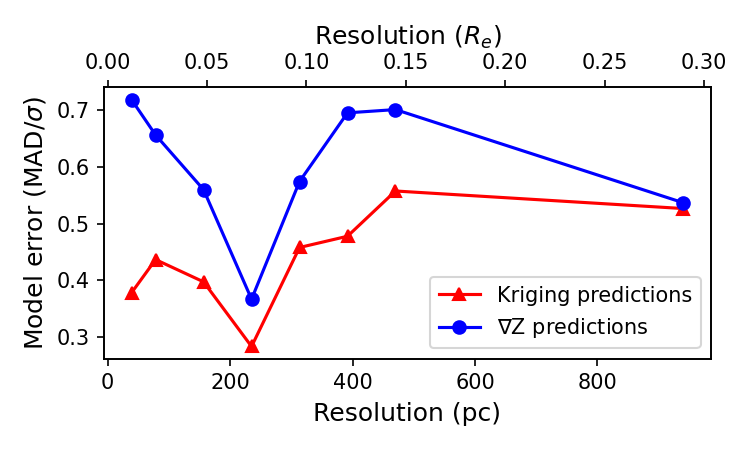}
    \caption{As in Figure \ref{fig:prediction_error_vs_res}, but using \ON\ as a metallicity diagnostic. We find that when $f=6$, a metallicity gradient alone provides an extremely good predictive model for the data. Even so, at all resolutions including when $f=6$, the linear gradient model is outperformed by a geostatistical model.}
    \label{fig:O3N2_pred_error}
\end{figure}

Finally, we show the results when the \ON\ diagnostic is used. In Figure \ref{fig:param_convergence_vs_res_O3N2}, we plot the median, 16\textsuperscript{th}, and 84\textsuperscript{th} percentiles in the posteriors of each parameter computed from a Monte-Carlo fit as a function of the resolution of the data input. We find that at native resolution, $\phi = 224^{ + 16 }_{ - 15 }$ pc, and $\gradZ$ is flatter than is found at coarser resolutions, consistent with the other two diagnostics discussed in this Appendix. Results are similar when $f=2$. However, an anomalous result is seen when $f=6$, and possibly at the two resolutions closest to this resolution. For this particular bin factor, using this diagnostic, the best-fit value of $\phi$ becomes very large, with large error bars: $\phi = 1.63^{ + 1.20 }_{ - 0.51 }$ kpc. This is larger than the value seen at any other resolution, including when $f=24$. Non-monotonic trends are also seen in the recovered values of $\gradZ$ and $\Zchar$ versus resolution when this diagnostic is used: at $f=6$, the value of $\gradZ$ reaches a minimum, and the (negatively correlated; $\rho = -0.57$) value of $\Zchar$ reaches a maximum. This indicates that for this particular resolution, with this particular diagnostic, the variance in the data can be preferentially described by a strong negative metallicity gradient, with no evidence for sub-kpc structures. 

Further evidence for this interpretation is seen in Figure \ref{fig:O3N2_pred_error}. When $f=6$, the median absolute difference between predicted metallicities from a linear gradient model and observed metallicities drops to $0.4\sigma$ -- significantly smaller than the WLS model error seen when any other diagnostic is used to analyse NGC 5236 at any resolution, and half of the value of MAD/$\sigma$ that is expected from Gaussian error. This indicates that for this combination of resolution and metallicity diagnostic, a linear gradient model does an excellent job of describing the metallicity variations within the data. Interestingly, the best-fit metallicity gradient found using a WLS method does not change significantly between $f=1$ and $f=10$ ($-0.012 \leq \gradZ \leq 0.09$ dex kpc$^{-1}$), indicating that this improvement in the performance of the model at this resolution is not caused by a more accurate recovery of the parameters, but by some difference in the properties of the data when this particular binning is used.

Even at $f=6$, when the data is very well-described by a simple linear trend, we find that our geostatistical model still outperforms the linear fit, with a lower MAD/$\sigma$ than is found even for the native resolution \ON\ metallicity map. We do not see any similar behaviour of the \ON\ diagnostic producing highly linear metallicity maps at particular resolutions for the other two galaxies investigated in this work. Because of this, we do not speculate as to what aspects of the \ON\ diagnostic may be responsible for this unusual behaviour. 

Clearly, there is much work left to be done in understanding how strong emission line based metallicity diagnostics calibrated on stacked, integrated galaxy spectra may be applied to sub-kpc regions of the ISM. Extending on the conclusions of \citet{Kewley+Ellison08}, we caution the reader that relying on individual strong emission line diagnostics may lead to erroneous conclusions about the metallicity structure of a galaxy. Following \citet{Kewley+19}, we recommend using a combination of multiple (three or more) strong emission line diagnostics using different line ratios to achieve a robust understanding of a galaxy's internal metallicity distribution.

\section{A primer on geostatistical methods}
\label{ap:extra_maths}
To capture details on the small-scale structure of the metallicity data around a large-scale metallicity trend, this work resorts to tools and techniques from \emph{geostatistics}, the field of mathematics and statistics that concerns itself with analysing stochastic processes that occur over a continuous spatial domain. This field is vast, having evolved over several decades since the seminal works of \citet{Matheron1954}, and we do not attempt to cover a comprehensive review here.\footnote{For those interested in geostatistics who want to read further, we recommend \citet{Wikle+19} as an introductory text, or \citet{Cressie93} for a comprehensive reference.} Rather, we introduce two important concepts that have been discussed extensively in the previous papers in this series: the \textit{semivariogram}, a tool for exploratory data analysis and separating correlated signals from uncorrelated noise, introduced for astronomical applications in \GeoGalsI; and \textit{universal kriging}, an interpolation method that takes as input a set of data points and a geostatistical model, and provides as output predictions of the value of a random field at a set of specified locations (discussed extensively in \GeoGalsII). Furthermore, we provide a comparison between the semivariogram, the two point correlation function and the power spectrum -- two popular tools in the analysis of spatial structure in astronomical datasets \citep[e.g.][]{PlanckPower, Li+21}.%, highlighting the advantages of the geostatistical approach for the analysis of galaxy data.

\subsection{The semivariogram and related methods}
\label{ap:SVGs}

To capture details on the small-scale structure of the metallicity data around the metallicity trend, we use the \textit{semivariogram}, a classic tool for exploratory data analysis from the geostatistical literature \citep{matheron1963}. The semivariogram is defined as:

\begin{equation}
    \gamma(h) = \frac12 \text{Var} \left( Z(\vec{x}) - Z(\vec{y}) \right),
\end{equation}
where the variance is computed over all pairs of points $\vec x$ and $\vec y$ for which $h - \delta_h/2  \leq | \vec x - \vec y| \leq h + \delta_h/2$. Here, $\delta_h$ is the size of bins used to compute the semivariogram, which should be chosen to be large enough so that enough pairs are present at each separation of interest that a variance can be reliably computed, but as small as possible to retain high spatial resolution \citep[e.g.][]{Cressie93, Journel+Huijbregts}.

Intuitively, the semivariogram shows how the variance between data points increases as the separation between them increases. Three informative parameters can be read from a semivariogram: the \textit{range}, which is the separation between data points at which the variance ceases to increase, indicative of the sizes of the largest substructures within the data; the \textit{sill}, which is the height of the semivariogram after it has flattened off, revealing the total variance in the data; and the \textit{nugget}, which is the height of the semivariogram at zero separation, equivalent to the amount of uncorrelated noise in the data. By comparing the height of the sill to the nugget, we can quantify the amount of correlated versus uncorrelated variance in the data, allowing the amplitude, in dex, of metallicity fluctuations to be inferred while ignoring the effects of uncorrelated instrumental and calibration errors -- see \GeoGalsI\ for further discussion and an illustration. %Furthermore, the overall shape of the semivariogram can be used to elucidate finer details on the nature of how metallicity is mixed within a galaxy.

The semivariogram is closely related to the two-point correlation function, $\xi(h)$, defined to be the Pearson correlation coefficient for the value of $Z$ at two points separated by a distance of $h$. Under the assumptions that the random field under investigation is stationary and isotropic, the two diagnostic functions are related by the following equation:

\begin{equation}
    \gamma(h) = \sigma^2\left(1-\xi(h)\right),
\end{equation}
where $\sigma^2$ is the overall variance of $Z$ throughout the field. Unlike the two-point correlation function, the semivariogram is not normalised; therefore, the semivariogram is sensitive to the amplitude of metallicity fluctuations, whereas the two-point correlation function is not. 

Another related metric often used in astronomy (and particularly cosmology) for determining the covariance structure of a random field is the power spectrum, which is calculated by taking the Fourier transform of a random field and then determining the square of the magnitude of each frequency component.
By the Wiener–Khinchin theorem, for particularly well-behaved random fields,
this function captures the same details that the semivariogram captures, but with units of spatial frequency rather than separation. However, the formal requirements on the data quality to use Fourier methods to reveal the covariance structure of the random field in this way are prohibitively strict: the random field must have no edges and no missing data, with periodic signals, and homoscedastic (constant variance) Gaussian error throughout \citep{Feigelson+Babu2012}. For analysis of the cosmic microwave background where these methods find their natural home, these requirements are readily satisfied. However, metallicity maps of galaxies are always finite fields; are heteroscedastic (not homoscedastic) due to galaxies being brighter (and therefore having higher S/N) in their central regions; and have large portions of missing data due to DIG contamination. We therefore recommend the semivariogram as a more natural approach for determining the covariance structure of galaxy properties such as metallicity.

\subsection{Universal kriging}
\label{ap:kriging}

One reason why geostatistical models are so powerful is that they are predictive. After a geostatistical model has been fit to noisy data, it can be used to predict what the true values of the variable in question are if the noise was not present. It can also be used to predict the values of variables at points where no measurement exists, or where measurements are unreliable. This can be valuable in the field of spatially-resolved metallicity analysis, where the metallicity found for many data points cannot be trusted due to low S/N or contamination from diffuse ionised gas (\GeoGalsII). 

\emph{Kriging} refers to a general family of inference techniques that provide unbiased, linear predictions for spatial data sets \citep{Stein1999} that are optimal in the sense that they minimise the mean square prediction error.
Many classes of kriging exist, including ordinary kriging (in which the data has a constant mean), simple kriging (in which the mean values of the data are not constant but follow a known trend), robust kriging (designed to be insensitive to outliers) and even Bayesian kriging (which provides predictions for data points by marginalising over a posterior of possible model parameters). More details on all of these methods can be found in \citet{Cressie93}. For our application, we use \emph{universal kriging}, in which the mean values of the data at each location are known to vary in space, but the best parameters to describe this trend are not known a priori, and must be learned from the data. 

The mathematics that powers universal kriging is not complicated, but several definitions are required to properly setup the mathematical framework that is needed to completely define this process.
We therefore refer the reader to \GeoGalsII\ for the necessary equations, and instead provide some intuition on how this kriging method works below.

In universal kriging, the value at an unknown location is inferred from (i) the best-fit value of the large-scale trend, and (ii) the (uncertain) values measured at all other data points. Based on the geostatistical model fit, the value of the unknown data point is expected to be strongly correlated with the values of data points nearby it, and weakly correlated with data points that are further away. In the model that we fit in this study, two data points separated by a distance of $\phi$ are expected to be positively correlated, with a correlation coefficient of $\rho = 0.37$ (Equation \ref{eq:matern_1/3}). This reflects the principle that things in nature that are close to each other tend to have similar properties (Tobler's First Law of Geography; \citealt{Tobler1970}). 

Unlike other prediction methods, in addition to producing a predicted value at each location, kriging methods also provide uncertainties associated with each prediction. The uncertainty in the universal kriging prediction is the sum of three components: (i) $\sigma^2$, the variance of the data, as fit by the geostatistical model (Equation \ref{eq:matern_1/3}); (ii) a term that represents the reduction in variance from the fact that the data point is known to be correlated to other nearby data points; and (iii) an additional factor that captures how the uncertainty of the parameters describing the global model increase the uncertainty of our estimate of the data value.

In the limiting case where the location at which a prediction is desired is very close to a measured data point, the value of the prediction becomes the measured value of that data point, and its uncertainty on the kriging prediction becomes the uncertainty on the measured value of that data point. In the other limiting case, when the prediction is made far from any known data, the predicted value of the data becomes the prediction provided by the large-scale model, and its uncertainty is equal to the uncertainty of the model, plus a factor that represents the degree to which data points are seen to be scattered around this mean trend. Between these two limiting cases, predictions come from some combination of local, uncertain measurements, and large-scale trends, with the geostatistical model of best fit controlling to what degree each of these factors
contribute to each prediction.

\section{Variance versus pixel size}
\label{ap:change_of_support}

In Section \ref{ssec:vs_res}, the total variance between \Hii spaxels was shown to depend on the resolution of the data, with lower resolution maps showing less intrinsic variance. Qualitatively, this result is expected. Lower resolution maps contain larger pixels, resulting in reported metallicities being averaged over a larger number of \Hii regions in each pixel. This causes all metallicities reported to be closer to the mean metallicity of the galaxy, reducing the variance of metallicity within the galaxy.

Quantitatively, however, this result is not so simple. How can we explain the ratios of the variance between the native resolution map and the maps with binning factor $f$? %\footnote{I'll be sure to have defined this in the main text, but a \textit{map with binning factor $f$} is made by binning blocks of $f\times f$ spaxels of the original resolution map. If the original data field has $N_x \times N_y$ spaxels then the map with $f=3$ has $N_x/3 \times N_y/3$ spaxels, each made of 9 native resolution spaxels.} 
Furthermore, if we had a finer resolution IFU observation of this galaxy, how much would we expect the variance in metallicity to increase? To answer these questions, we must consider the correlation structure of the underlying metallicity random field, $Z(\vec{x})$, and the size of the pixels used to observe the random field (known in the geostatistical literature as the \textit{support} of the data).

To understand how the visible features of a random field change as the support is changed, we follow \citet{Gelfand+10}. For each map with binning factor $f$, the true value of the metallicity at a point, $Z(\vec{x})$, is not observed. Instead, only the average metallicity over an area $A_f$ can be seen:

\begin{equation}
    Z_{f}(\vec{x}_0) = \frac{1}{|A_{f}|} \int_{A_{f0}} Z(\vec{s})d\vec{s}.
\end{equation}

Here, $A_{f0}$ is an areal patch centred on $\vec{x}_0$, with area $|A_{f}|$. For our purposes, we model each area $A_{f0}$ as a square patch of sky centred on $\vec{x}_0$ with side length equal to $f \times 39$ pc, recalling that $39$ pc is the native resolution of our observations of NGC 5236.

The covariance of $Z_{f}(\vec{x})$ can be calculated from the covariance of $Z(\vec{x})$. We assume that the covariance between the metallicity at two points depends only on their separation: that is, that there exists a function $C(\vec{x} - \vec{y})$ such that Cov$(Z(\vec{x}), Z(\vec{y})) = C(\vec{x} - \vec{y})$ for all points $\vec{x}, \vec{y}$ in the galaxy. In this case, the covariance between two observations averaged over areas $A_{f1}$ and $A_{f2}$ centred on two point $\vec{x}_1, \vec{x}_2$ is given by:

\begin{equation}
    \text{Cov}(Z_{f}(\vec{x}_1),Z_{f}(\vec{x}_2))  = \frac{1}{|A_f|^2} \int_{A_{f1}}\int_{A_{f2}} C(\vec{s}_1 - \vec{s}_2)d\vec{s}_2 d\vec{s}_1.
\end{equation}

If the metallicity of every spaxel was independent, we would expect the ratio of the variance in metallicity between the native resolution metallicity map and each map with degraded resolution to decrease as $1/N$, where $N=f^2$ is the number of native resolution spaxels contained within each degraded resolution spaxel. However, such a model predicts a much steeper decline in variance as the binning factor $f$ increases than is seen in our data. Furthermore, as the resolution of the data is increased, under this model, we would expect the observed variance continue to increase indefinitely. This would be worrisome, as it would imply that the variance in metallicity seen within a galaxy at any finite resolution could never reflect the true intrinsic metallicity variability within a galaxy.

Instead, if we assume that the correlation between nearby values is approximately exponential, we get the following relation: $C(\vec{x} + \vec{y}) \approx \frac{1}{\sigma^2}C(\vec{x})C(\vec{y})$. Using this, we can relate the covariance of the averaged metallicities within each spaxel to the covariance of the intrinsic metallicity distribution by changing the variable $\vec{s}_2$ to $\vec{s}' = \vec{s}_2 - \vec{x}_2 + \vec{x}_1$:

\begin{eqnarray}
\text{Cov}(Z_{f}(\vec{x}_1),Z_{f}(\vec{x}_2)) = \frac{1}{|A_f|^2} \int_{A_{f1}}\int_{A_{f2}} C(\vec{s}_1 - \vec{s}_2)d\vec{s}_2 d\vec{s}_1 \nonumber \\
= \frac{1}{|A_f|^2} \int_{A_{f1}}\int_{A_{f1}} C(\vec{s}_1 - \vec{x}_2 + \vec{x}_1 - \vec{s'})d\vec{s}' d\vec{s}_1 \nonumber \\
= C(\vec{x}_1 -\vec{x}_2) \frac{1}{|A_f|^2 \sigma^2} \int_{A_{f1}}\int_{A_{f1}} C(\vec{s}_1 -\vec{s}')d\vec{s}' d\vec{s}_1. \label{eq:reduction_in_cov}
\end{eqnarray}

This shows that the covariance between the spatially-averaged quantities $Z_{f}(\vec{x}_1)$ and $Z_{f}(\vec{x}_2)$ is, to first order, the same as the covariance between the intrinsic quantities $Z(\vec{x}_1)$ and $Z(\vec{x}_2)$, up to a factor of $K := \frac{1}{|A_f|^2 \sigma^2} \int_{A_{f1}}\int_{A_{f1}} C(\vec{s}_1 -\vec{s}')d\vec{s}' d\vec{s}_1$. This means that for a given model of the spatial correlation between data points, we can calculate this reduction in variance, $K$, as a function of pixel sizes.

\begin{figure}
    \centering
    \includegraphics[width=0.49\textwidth]{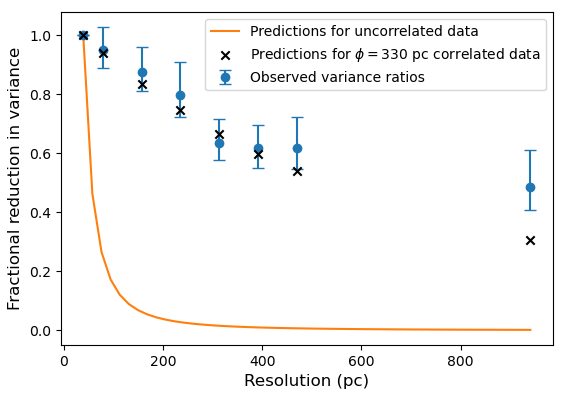}
    \caption{Blue points show the ratio of the height of the semivariograms pictured in Figure \ref{fig:svg_vs_res} compared to the native resolution map with $f=1$. Black crosses show the analytic expected reduction in variance when data is spatially correlated with a correlation scale length of $\phi=330$pc, following an exponential covariance structure. The orange line shows the naive reduction in variance expected when spatial correlations are not accounted for.} 
    \label{fig:reduction_of_variance}
\end{figure}

We show the results of this calculation in Figure \ref{fig:reduction_of_variance} for NGC 5236 with our fiducial metallicity diagnostic, assuming an exponential covariance function with a correlation scale length of $\phi=330$ pc (the best fit value calculated from the native resolution data; Section \ref{ssec:vs_res}). We see that the reduction in variance as $f$ is increased seen in the semivariograms displayed in Figure \ref{fig:svg_vs_res} is consistent with the analytical predictions of this model. We also see that the reduction in variance is far less than would be assumed in a model in which no spatial correlations are accounted for.

Under this framework, we can also determine how much variance is lost when our native resolution map is used. If our resolution was increased to $1$ pc (for example, due to followup observations from ELT class instruments), our variance would only increase by a factor of $1.06$, far smaller than the $39$-fold increase that would be naively expected if spatial correlations were not accounted for. Because of this, we can be justified in saying that the variances we report for the small-scale metallicity structure in this galaxy have physical meaning, and are not merely functions of the resolution at which the galaxy is viewed.

%We note that this change-of-support argument applications. When generalised to allow regions to have different sizes, it can be used to link the underlying metallicity field to metallicity maps produced by binning many pixels \citep[e.g.][]{Papaderos+02, Cappellari+Copin03, Esposina-Ponce+20,Li+22},  which aim to have uniform S/N at the cost of non-uniform pixel sizes.

\label{lastpage}
\end{document}